\documentclass[11pt,pra,aps,epsfig,psfig,multicols,showpacs,tightenlines,floatfix]{revtex4}
\usepackage{float}
\usepackage{graphics,bm}
\usepackage{graphicx}
\usepackage{amsmath, amssymb, graphics}
\allowdisplaybreaks
\usepackage{dcolumn}
\usepackage{bm}
\usepackage{multirow,array}
\usepackage{graphicx}
\usepackage{amsmath}
\usepackage{float}
\usepackage{xcolor}
\usepackage{empheq}

\newcommand{\beq}{\begin{equation}}
\newcommand{\eeq}{\end{equation}}
\newcommand{\bqa}{\begin{eqnarray}}
\newcommand{\eqa}{\end{eqnarray}}
\def\gsim{\mathrel {\vcenter {\baselineskip 0pt \kern 0pt
\hbox{$>$} \kern 0pt \hbox{$\sim$} }}}
\def\lsim{\mathrel {\vcenter {\baselineskip 0pt \kern 0pt
\hbox{$<$} \kern 0pt \hbox{$\sim$} }}}

\protect

\begin{document}

 \title{Reflectionless potentials and resonant scattering of flat-top and thin-top solitons}

\date{}

\author{L. Al Sakkaf  and U. Al Khawaja}

\address{
\it Department of Physics, United Arab Emirates University, P.O.
Box 15551, Al-Ain, United Arab Emirates}


\begin{abstract} \noindent
We identify a class of potentials for which the scattering of
flat-top solitons and thin-top solitons of the nonlinear
Schr\"odinger equation with dual nonlinearity can be reflectionless.
The scattering is characterized by sharp resonances between regimes
of full transmission and full quantum reflection. Perturbative
expansion in terms of the magnitude of radiation losses leads to the
general form of reflectionless potentials. Simulating the scattering
of flat-top solitons and thin-top solitons confirms the
reflectionless feature of these potentials.
\end{abstract}

\maketitle

\section{Introduction}
\label{intsec} The unique features of solitons are exhibited most
clearly upon their scattering by external potentials. The so-called
reflectionless scattering is one such striking example. Scattering
of solitons by specific types of potentials is characterized by the
complete absence of radiation and by preserving the integrity of
solitons after scattering \cite{goodman}. Moreover, the interesting
phenomena of quantum reflection and resonant scattering are usually
studied with reflectionless potentials \cite{lee,ernst}. A
well-known and well-studied example is the scattering of bright
solitons of the nonlinear Schr\"odinger equation (NLSE) by the
reflectionless P\"oschl-Teller potential well \cite{goodman,lee}.

On the fundamental level, resonant scattering by reflectionless
potentials reveals the spectrum of the potential's bound states
\cite{LU}. Reflectionless scattering is also important for
applications in optical data processing where optical devises may be
designed to simulate certain functions such as switching, routing,
unidirectional flow, and logic gating
\cite{hasegawa,asad,usama1,usama2}.

Stimulated by the recent experimental realization of the  flat-top
soliton \cite{exp1,exp2}, which is a solution of the NLSE with dual
nonlinearity, interest in such kind of solitonic excitations has
been growing
\cite{cheiney,bottcher,luo,khare,usama3,debnath,petrov,chen,pathak}.
However, and to the best of our knowledge, the scattering properties
of flat-top solitons are scarcely studied in the literature
\cite{zeng,umarov1,umarov2}. The present work attempts to answer three
related fundamental questions: 1) Can quantum reflection occur for
flat-top solitons? 2) Do reflectionless potentials exist for
flat-top solitons? 3) Do flat-top solitons exhibit resonant
scattering, as in bright solitons of NLSE? The rationale behind
posing these questions is that flat-top solitons can be very wide,
and thus it is not obvious that such a wide object may scatter off
potentials without splitting or forming radiation. Nonetheless, the
answer we found to all of these questions is `yes'.

We consider the NLSE with dual nonlinearity and then show that it
supports a class of solutions that comprises a spectrum of flat-top
solitons, bright solitons, kink solitons, and another type which we
denote as `thin-top' solitons. The spectrum of flat-top solitons is
bounded by bright solitons, with vanishing width of its top, and kink
solitons, with diverging width of its top.
We construct a new type of reflectionless
potential using the flat-top soliton and then study its scattering
properties. The form of the reflectionless potential is derived
using an ansatz for the scattered wave in the form of a localized
soliton part and a small extended radiation part. A perturbative
expansion in the small magnitude of the radiation part leads to the
general form of reflectionless potential. From the spectrum of this
class of solutions, a corresponding spectrum of reflectionless
potentials is obtained. Simulating the scattering of flat-top and
thin-top solitons by these reflectionless potentials, the
reflectionless feature of the potentials will then be revealed. In
addition, our simulations show that flat-top and thin-top solitons
do indeed exhibit quantum reflection and resonant scattering. It
should be noted that simulating flat-top solitons turns out to be
numerically demanding when the soliton is very wide. Our numerical
code, based on power series expansion \cite{LUcode}, is shown to
efficiently simulate the widest flat-top soliton possible by the
machine precision.

The rest of the paper is organized as follows. In Section
\ref{solssec}, we present the exact solution of the NLSE with dual
nonlinearity and discuss its properties. In Section \ref{refsec}, we
derive the reflectionless potential. In Section \ref{scatsec}, we
perform numerical simulations that confirm the reflectionless
property of the potentials. In Section \ref{concsec}, we end with
our main conclusions.

\section{Family of solutions}
\label{solssec}

\begin{figure}[H]\centering
    \includegraphics[width=9cm,clip]{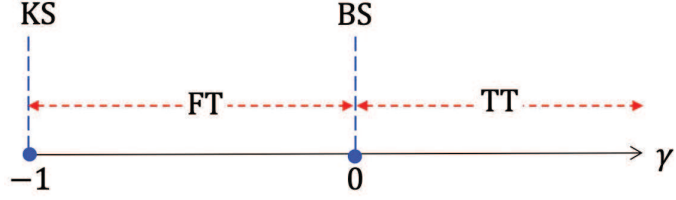}
        \caption{Schematic diagram showing the spectrum of solutions in terms of $\gamma$. The four solutions, bright soliton (BS),  kink soliton (KS), flat-top soliton (FT), and thin-top soliton (TT) are given by (\ref{solbs}), (\ref{solks2}), (\ref{solft}), (\ref{soltt}), respectively.}
    \label{fig1}
\end{figure}

\begin{figure}[H]\centering
    \includegraphics[width=8.3cm,clip]{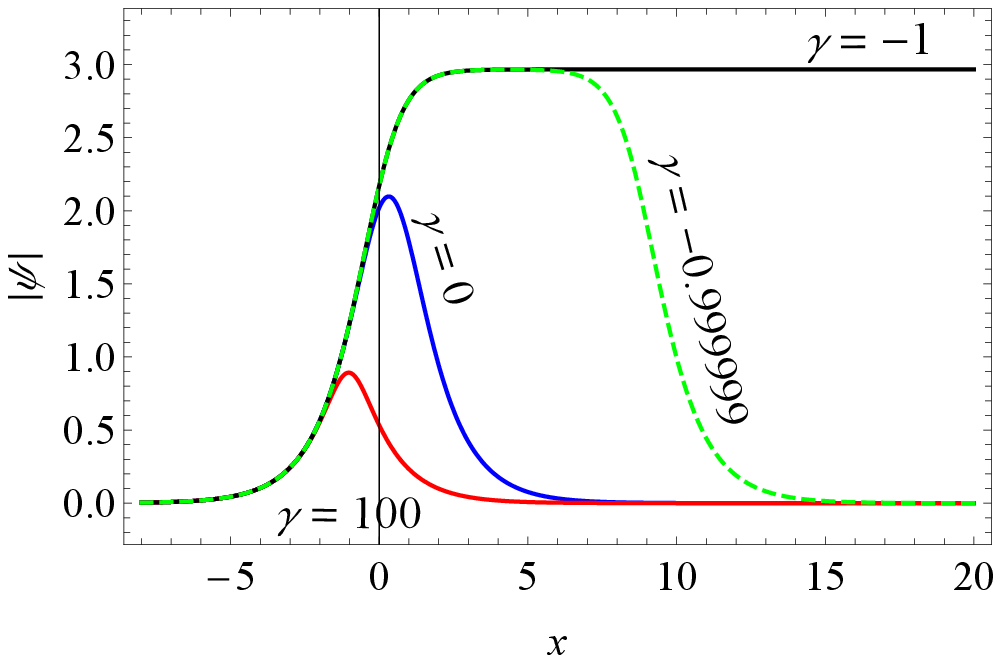}
        \vspace*{0.6cm}
    \begin{picture}(5,5)(5,5)
    \put(-200,144) {\color{black}{{\fcolorbox{white}{white}{\textbf{(a)}}}}}
    \end{picture}
    \includegraphics[width=8.5cm,clip]{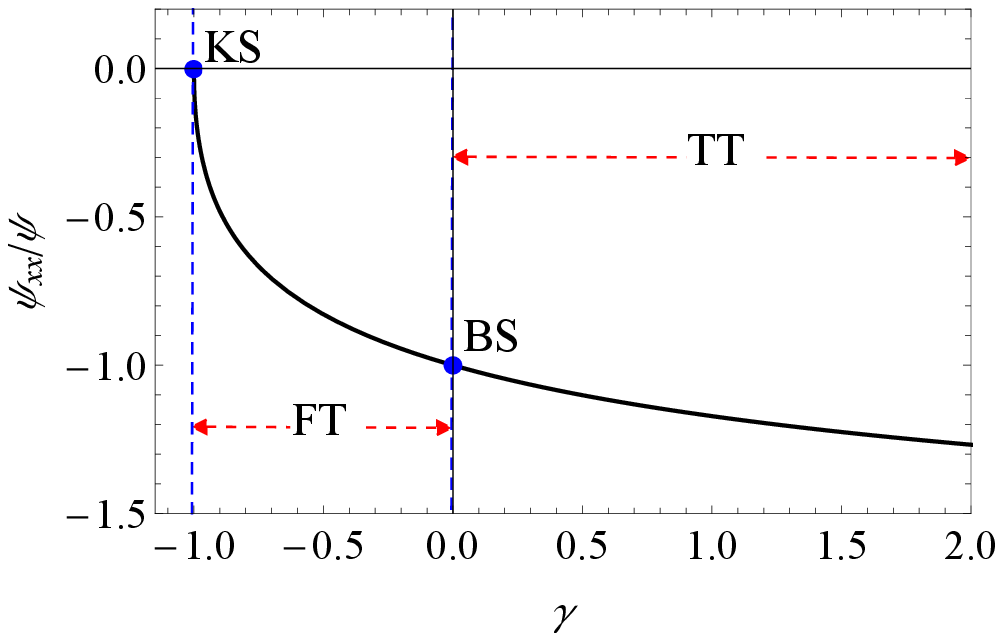}
    \begin{picture}(5,5)(5,5)
    \put(-30,144) {\color{black}{{\fcolorbox{white}{white}{\textbf{(b)}}}}}
    \end{picture}
    \caption{Family of exact solutions to Eq. (\ref{dnlse}). (a) Profiles of the four solutions (\ref{solbs}), (\ref{solks2}), (\ref{solft}), (\ref{soltt}). (b) Curvatures of the four solutions at $x=t=0$ with $u_0=g_1=1$.   Parameters used in (a): $u_0=2.2$, $v_0=0.0$, $g_1=3.0$, $g_2=1.0$,
    $g_3=\gamma\, g_{30}$, $g_{30}=3g_2^2/16u_0$, and $x_0$ is given by (\ref{x0eq}).}
    \label{fig2}
\end{figure}

\begin{figure}[H]\centering
    \includegraphics[width=8.5cm,clip]{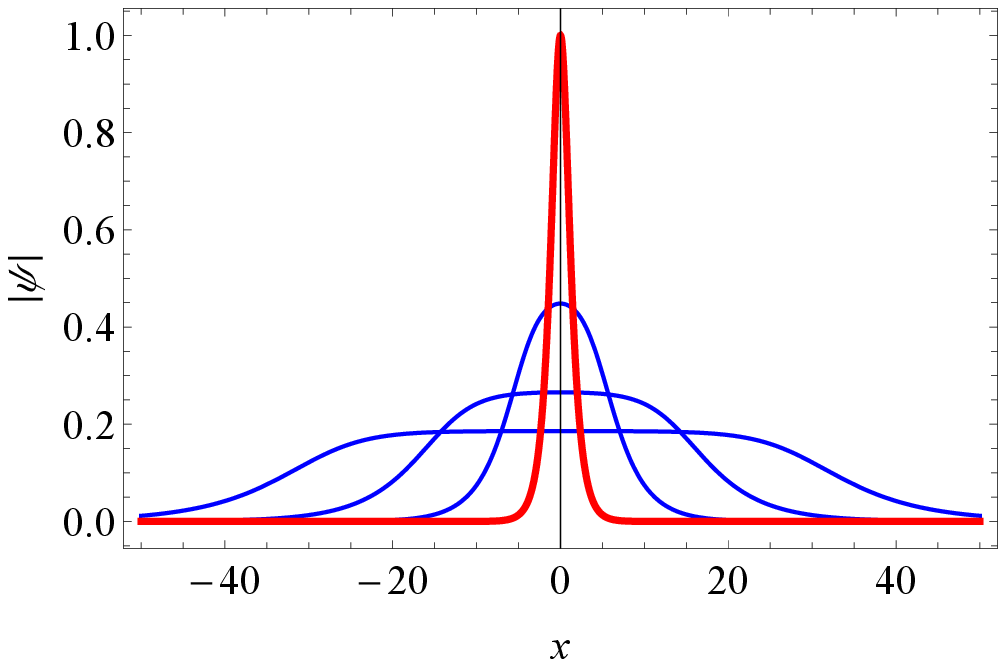}
        \vspace*{1.2cm}
    \begin{picture}(5,5)(5,5)
    \put(-30,144) {\color{black}{{\fcolorbox{white}{white}{\textbf{(a)}}}}}
    \put(-152,170) {\color{black}{{\fcolorbox{white}{white}{\textbf{Flat-Top solitons}}}}}
    \end{picture}
    \includegraphics[width=8.28cm,clip]{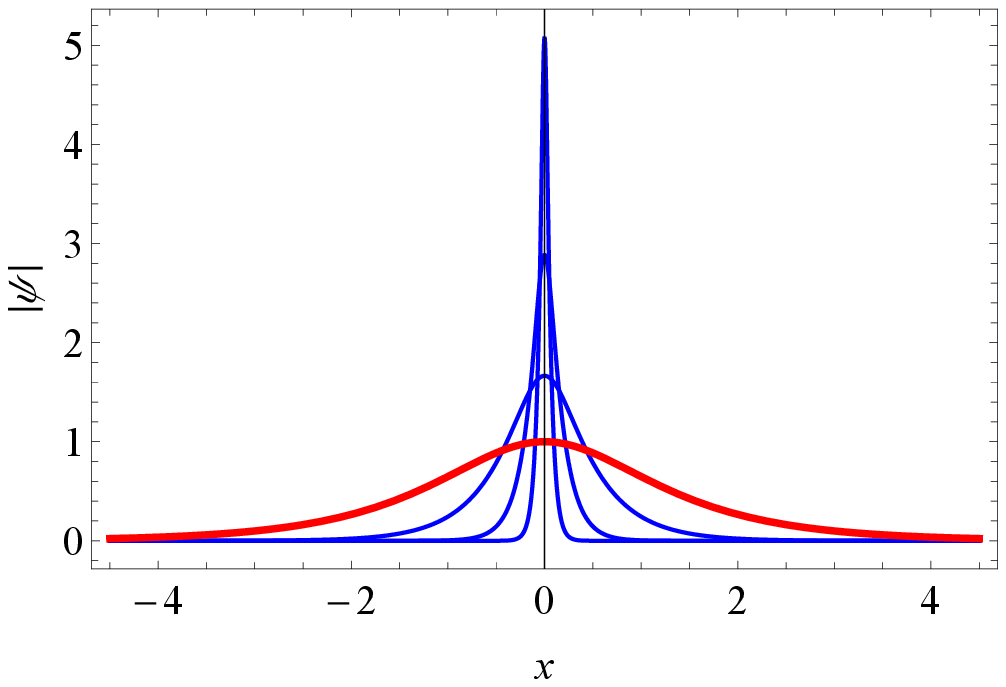}
    \begin{picture}(5,5)(5,5)
    \put(-30,144) {\color{black}{{\fcolorbox{white}{white}{\textbf{(b)}}}}}
        \put(-152,170) {\color{black}{{\fcolorbox{white}{white}{\textbf{Thin-Top solitons}}}}}
    \end{picture}
    \caption{(a) Flat-top solitons, given by (\ref{solft}),
    with $\gamma=\{-0.99, -0.9999, -0.999999\}$,
    such that the last value corresponds to the widest soliton, and (b) thin-top solitons,
    given by (\ref{soltt}), for $\gamma=\{9,99,999\}$, such that the last value corresponds to the
    thinnest soliton. Thick (red) curves correspond to the bright soliton solution for $\gamma=0$.
    Parameters used in both (a) and (b): $u_0=2.2$, $v_0=0.0$, $g_1=3.0$, $g_2=1.0$, $g_3=\gamma\,
    g_{30}$, $g_{30}=3g_2^2/16u_0$.}
    \label{fig3}
\end{figure}

The dynamics of so-called flat-top soliton is described by the
following NLSE with dual nonlinearity
\begin{equation}
i\frac{\partial}{\partial
t}\psi(x,t)=-g_1\,\frac{\partial^2}{\partial
x^2}\psi(x,t)-g_2|\psi(x,t)|^n\psi(x,t)-g_3|\psi(x,t)|^{2n}\psi(x,t)+V(x)\,\psi(x,t)
\label{dnlse},
\end{equation}
where $\psi(x,t)$ is generally a complex field, $V(x)$ is an
external potential, $g_1$ characterizes the strength of dispersion,
$g_2$ and $g_3$ characterize the strengths of the two
nonlinearities, and $n$ is an integer. For $n=2$, which we will
assume for the rest of this paper, the typical NLSE with cubic and
quintic nonlinearities will be retrieved, where the cubic term
corresponds to the Kerr nonlinearity in the context of optical
solitons and Hartree-Fock interatomic interaction in the case of
Bose-Einstein condensates \cite{pethick}. The flat-top soliton
solution belongs to a more general class of stationary solutions
which can be written in the form \cite{usamaNbahlouli, ourbook}
\begin{equation}
\psi(x,t)=\sqrt{\frac{2u_0}{g_2\sqrt{1+\gamma}}}\,\frac{1}{\sqrt{\frac{1-\sqrt{1+\gamma}}{2\sqrt{1+\gamma}}+{\rm
cosh}^2\left[\sqrt{\frac{u_0}{g_1}}(x-x_0-v_0t)\right]}}\,e^{i\phi(x,t)}
\label{sol1},
\end{equation}
where $u_0$, $x_0$, and $v_0$ are arbitrary parameters corresponding
to the amplitude, peak position, and speed of the soliton,
respectively, $\gamma=g_3/g_{30}$, where $g_{30}=3g_2^2/16u_0$, and
$\phi(x,t)=u_0\,t+v_0[2(x-x_0)-v_0\,t]/4g_1$. It should be noted
that, although we denote this solution as stationary, we still allow
for the centre-of-mass motion in order to study its scattering with
potentials below. The profile is thus truly stationary only in a
frame of reference moving with the soliton. For stationary
solutions, the profile should be real which requires $\gamma\ge-1$.
This general expression defines a spectrum of solutions containing
four different nontrivial solutions, namely: bright soliton (BS),
kink soliton (KS), flat-top soliton (FT), and {\it thin-top} soliton
(TT). We termed the last solution as such since it turned out that
its peak width is thinner than that of the bright soliton, which is
the opposite case of flat-top solitons. We believe this type of
soliton is not identified or studied previously in the literature.
The whole spectrum of solutions can be scanned in terms of $\gamma$
values, as shown schematically in Fig. \ref{fig1}.
We list here the four different solutions. While some of these solutions are known, we include them for completeness.\\\\
{\bf I. Bright soliton, $\gamma=0$}:
\begin{equation}
\psi(x,t)=\sqrt{\frac{2u_0}{g_2}}\,{\rm
sech}\left[\sqrt{\frac{u_0}{g_1}}(x-x_0-v_0t)\right] \,e^{i\phi(x,t)}
\label{solbs},
\end{equation}
{\bf II. Kink soliton, $\gamma=-1$}:\\
Taking the limit $\gamma\rightarrow-1$ in Eq.~(\ref{sol1}), the
solution takes the form of the following constant wave (CW)
\begin{equation}
\psi(x,t)=2\,\sqrt{\frac{u_0}{g_2}}\,e^{i\phi(x,t)} \label{solks1}.
\end{equation}
Since $x_0$ is an arbitrary parameter, we set $x_0$ as
\begin{equation}
x_0=\sqrt{\frac{g_1}{16u_0}}\,\ln{\left[\frac{g_1}{g_2^2(1+\gamma)}\right]}
\label{x0eq},
\end{equation}
which corresponds to shifting the origin such that the left side of
the flat-top solution is always at $x=0$ for all values of $\gamma$.
In this case the limit $\gamma\rightarrow-1$ leads to the following
kink solution
\begin{equation}
\psi(x,t)=2\,\sqrt{\frac{u_0}{g_2}}\,
\frac{1}{\sqrt{1+\sqrt{\frac{g_1}{4g_2^2}}\,\exp{\left(-2\sqrt{\frac{u_0}{g_1}}\,x\right)}}}
\,e^{i\phi(x,t)} \label{solks2}.
\end{equation}
{\bf III. Flat-top soliton, $-1<\gamma<0$}: For this range of
$\gamma$, it will be convenient to express the solution using the
transformation $\gamma=-\cos^2(\theta)$, where $0<\theta<\pi/2$
\begin{equation}
\psi(x,t)=2\,\sqrt{\frac{u_0}{g_2}}\,
\frac{1}{\sqrt{1+\sin(\theta)\,\cosh\left[\sqrt{\frac{4u_0}{g_1}}(x-x_0-v_0t)\right]}}
\,e^{i\phi(x,t)} \label{solft}.
\end{equation}
{\bf IV. { \emph{\textbf{Thin-top}}} soliton, $\gamma>0$}:\\
In this case, the solution is expressed using the transformation
$\gamma=\sinh^2(\theta)$, where $\theta\ne0$
\begin{equation}
\psi(x,t)=2\,\sqrt{\frac{u_0}{g_2}}\,
\frac{1}{\sqrt{1+\cosh(\theta)\,\cosh\left[\sqrt{\frac{4u_0}{g_1}}(x-x_0-v_0t)\right]}}
\,e^{i\phi(x,t)} \label{soltt}.
\end{equation}
In Fig. \ref{fig2}(a),  we plot these four solutions for the same set of parameters. They differ only by the value of $\gamma$ and hence their norm.
The central curvature of flat-top soliton is smaller than that of
bright soliton. In contrast, the central curvature of the thin-top
soliton is larger than that of bright soliton.  In Fig.
\ref{fig2}(b), we plot $(1/|\psi|)d^2|\psi|/dx^2$ at $x=0$. The
figure shows that the curvature of flat-top solitons is always
smaller than that of the bright soliton, and the curvature of
thin-top solitons is always larger than the curvature of the bright
soliton. The fundamental difference between flat-top soliton and
thin-top soliton can be also clearly seen upon comparing them when
they have the same norm. Normalising the general solution as
\begin{equation}
n_0=\int_{-\infty}^{\infty}|\psi(x,t)|^2dx=\frac{8g_1}{g_2}\,\sqrt{\frac{u_0}{g_1\,\gamma}}\,\tan^{-1}\left(\frac{\sqrt{1+\gamma}-1}{\sqrt{\gamma}}\right)\label{n0eq},
\end{equation}
and then solving for $u_0$,
\begin{equation}
u_0=\frac{g_2^2n_0^2\gamma}{64g_1\,\left[\tan^{-1}\left(\frac{\sqrt{1+\gamma}-1}{\sqrt{\gamma}}\right)\right]^2}
\label{u0eq},
\end{equation}
the general solution (\ref{sol1}), and hence the special solutions (\ref{solbs}), (\ref{solks1}), (\ref{solks2}), (\ref{solft}), (\ref{soltt}), will be normalised to $n_0$. In Fig. \ref{fig3}, we plot
the normalised flat-top and thin-top solitons together with the bright soliton for the purpose of comparison.
The figure shows that flat-top solitons are indeed always wider than
the bright soliton and they become wider for values of $\gamma$
approaching $-1$, and thin-top solitons are always thinner than the
bright soliton and they become thinner as $\gamma$ approaches
$\infty$.

Finally, a statement about the stability of these solitons is in
order. According to the Vakhitov-Kolokolov (VK) stability criterion
\cite{vakhitov}, the soliton with profile
$\psi(x,t)=\phi(x)e^{-i\omega\,t}$, will be stable if the condition
$\partial n_0/\partial\omega<0$ is satisfied. Applying this
criterion on solution (\ref{sol1}), where $\omega=-u_0$, we get
\begin{equation}
\frac{\partial}{\partial\omega}n_0=-\frac{4}{g_2}\,\sqrt{-\frac{g_1}{\omega}}\,\frac{\tan^{-1}\left(\frac{\sqrt{1+\gamma}-1}{\sqrt{\gamma}}\right)}{\sqrt{\gamma}}
\label{stabeq}.
\end{equation}
For $g_1,\,g_2>0$ and $\omega<0$, this expression is negative for $\gamma\ge-1$, which indicates that all types of solitons mentioned above are stable according to the VK criterion.

In the next section, we derive the reflectionless potentials
corresponding to the above solutions.

\section{Reflectionless potentials}
\label{refsec} Reflectionless scattering is defined by the absence
of radiation. The scattering outcome is then assumed to be in the
form of a localized solitonic part that is immersed  in a weak
oscillatory part, which accounts for radiation. A perturbative
approach will then lead to the specific form of potential for which
radiation is vanishingly small. The ansatz for the scattering
outcome is written as
\begin{equation}
\psi_{out}(x,t)=\left[\psi_0(x)+\psi_1(x,t)\right]\,e^{i\lambda t}
\label{ansatz},
\end{equation}
where $\psi_0(x)\,e^{i\lambda t}$ is the stationary solitonic part
and $\psi_1(x,t)$ is the small radiation part, such that $|\psi_1|
\ll |\psi_0|$. The solitonic part corresponds to either reflected or
transmitted soliton long after the scattering event such that the
effect of the finite-range potential is absent. Substituting this
ansatz in (\ref{dnlse}), the zeroth order vanishes assuming that
$\psi_0(x)$ is a solution of the time-independent version of
(\ref{dnlse}), namely
\begin{equation}
g_1\,\psi_0^{\prime\prime}(x)+g_2\,\psi_0^3(x)+g_3\,\psi_0^5(x)-\left(V(x)+\lambda\right)\,\psi_0(x)=0
\label{zeroth}.
\end{equation}
Since we consider a situation where the soliton after scattering is
far from the potential $V(x)$, the soliton solution $\psi_0(x)$ may
be taken as the solution of the fundamental version of
(\ref{dnlse}), i.e., Eq.~(\ref{zeroth}) with $V(x)=0$. The linear
order gives
\begin{equation}
i\frac{\partial}{\partial
	t}\psi_{1}(x,t)+g_1\frac{\partial^2}{\partial
	x^2}\psi_1(x,t)+\left[g_2\,\psi_0^2(x)+g_3\,\psi_0^4(x)-(V(x)+\lambda)\right]\,\psi_1(x,t)+\left(2g_3\,\psi_0^4+g_2\,\psi_0^2\right){\psi_1}^*(x,t)=0
\label{linear}.\end{equation} In terms of the real and imaginary
parts of $\psi_1(x,t)$, defined by
$\psi_1(x,t)=\psi_{1r}(x,t)+i\,\psi_{1i}(x,t)$, the real and
imaginary parts of the last equation give
\begin{equation}
g_1\,\frac{\partial^2}{\partial
x^2}\psi_{1r}-\,\frac{\partial}{\partial
t}\psi_{1i}+\left[5g_3\,\psi_0^4+3g_2\,\psi_0^2-(V(x)+\lambda)\right]\psi_{1r}=0
\label{realeq},
\end{equation}
\begin{equation}
g_1\,\frac{\partial^2}{\partial
x^2}\psi_{1i}+\frac{\partial}{\partial
t}\psi_{1r}+\left[g_3\,\psi_0^4+g_2\,\psi_0^2-(V(x)+\lambda)\right]\psi_{1i}=0
\label{imeq}.
\end{equation}
We set the oscillatory perturbations in the form
\begin{equation}
\psi_{1r}(x,t)=u\cos(k\,x+\omega t),
\end{equation}
and
\begin{equation}
\psi_{1i}(x,t)=v\sin(k\,x+\omega t),
\end{equation}
where $u$ and $v$ are arbitrary amplitudes. Substituting back in
Eqs.~(\ref{realeq}) and (\ref{imeq}), a nontrivial solution requires
\begin{equation}
V(x)=-\left[g_1\,k^2+\lambda-2g_2\,\psi_0^2(x)-3g_3\,\psi_0^4(x)\pm
\sqrt{\omega^2+g_2^2\,\psi_0^4(x)+4g_2\,g_3\,\psi_0^6(x)+4g_3^2\,\psi_0^8(x)}\right]
\label{refpot}.
\end{equation}
In the long-wavelength limit, $k,\,\omega\rightarrow0$, the last
equation reduces to
\begin{equation}
V(x)=g_2\,\psi_0^2(x)+g_3\,\psi_0^4(x) \label{refpot1},
\end{equation}
or
\begin{equation}
V(x)=3g_2\,\psi_0^2(x)+5g_3\,\psi_0^4(x) \label{refpot2},
\end{equation}
where we have also set $\lambda=0$ since it corresponds to a constant
shift in energy.  In the absence of the quintic nonlinearity,
the above result should lead to the well-known case of
P\"oschl-Teller reflectionless potential, namely
\begin{equation}
V(x)=-2\,u_0\,{\rm
sech^2\left(\sqrt{{\it u}_0/g_1}{\it x}\right)}\label{ptpot}.
\end{equation}
Using the bright soliton solution $\psi_0(x)=\sqrt{2\,{\it
u}_0/g_2}\,{\rm sech}(\sqrt{u_0/g_1}{\it x})$ in Eqs.
(\ref{refpot1}) and (\ref{refpot2}) gives $V(x)=2\,u_0\,{\rm
sech}^2\left(\sqrt{{\it u}_0/g_1}{\it x}\right)$, and
$V(x)=6\,u_0\,{\rm sech}^2\left(\sqrt{{\it u}_0/g_1}{\it
x}\right)$, respectively. We conjecture that what matters for the reflectionless
feature is the structure of the potential such that a prefactor will
not affect this property. This is based on our numerical
simulations. Our perturbative calculation, in this case, accounts for
the structure of the reflectionless potential but does not explain
the freedom in the parameters. With this interpretation, the
P\"oschl-Teller potential is obtained from our results by
multiplying the potential expressions we obtained by an arbitrary
overall parameter that can be adjusted to get the right sign and
magnitude.  Indeed, our
simulations show that the reflectionless property is not restricted
to the parameters of the model we are solving, namely Eq.
(\ref{dnlse}). The potential can thus be constructed by another
solution that has the same shape but with different parameters,
namely $\psi_{0p}(x)=\sqrt{2\,{\it u}_{0p}/g_{2p}}\,{\rm
sech}(\sqrt{{\it u}_{0p}/g_{1p}}{\it x})$, where $g_{1p}$, $g_{2p}$,
and $u_{0p}$ are arbitrary parameters that may be different than
$g_1$, $g_2$, and $u_0$.

\begin{figure}[H]\centering
    \includegraphics[width=9cm,clip]{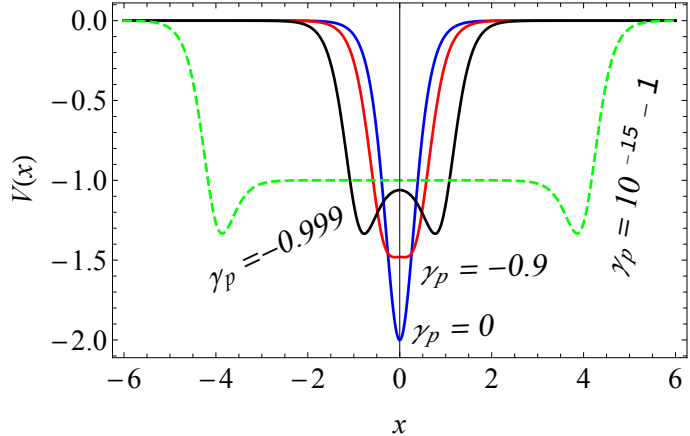}

    \caption{Potential profiles of (\ref{refpt2}) with different values of $\gamma_p$. Parameters used:  $u_{0p}=1$, $g_{1p}=0.2$, $V_0=1$.}
    \label{fig4}
\end{figure}

Similarly, a reflectionless potential for flat-top and thin-top
solitons is obtained by extracting $\psi_0(x)$ from (\ref{sol1}) and
substituting in (\ref{refpot}), which is then simplified as

\begin{equation}
V(x)=-\frac{2u_{0}}{\sqrt{1+\gamma}}\,
\frac{\frac{3\gamma}{8\sqrt{1+\gamma}}
	+\frac{1-\sqrt{1+\gamma}}{2\sqrt{1+\gamma}}
	+\cosh^2\left(\sqrt{\frac{u_{0}}{g_{1}}}x\right)}
{\left[\frac{1-\sqrt{1+\gamma}}{2\sqrt{1+\gamma}}
	+\cosh^2\left(\sqrt{\frac{u_{0}}{g_{1}}}x\right)\right]^2}
\label{refpt2n},
\end{equation}
and re-expressed with new free parameters  ${u_{0}}_{p}$, $\gamma_p$, ${g_{1}}_p$, and an overall arbitrary parameter $V_0$ as 

\begin{equation}
V(x)=-V_{0}\,\frac{2u_{0p}}{\sqrt{1+\gamma_p}}\,
\frac{\frac{3\gamma_p}{8\sqrt{1+\gamma_p}}
+\frac{1-\sqrt{1+\gamma_p}}{2\sqrt{1+\gamma_p}}
+\cosh^2\left(\sqrt{\frac{u_{0p}}{g_{1p}}}x\right)}
{\left[\frac{1-\sqrt{1+\gamma_p}}{2\sqrt{1+\gamma_p}}
+\cosh^2\left(\sqrt{\frac{u_{0p}}{g_{1p}}}x\right)\right]^2}
\label{refpt2}.
\end{equation}
Here, $\gamma_p>-1$ is an arbitrary parameter and we have added $V_{0}$ to be able to change the depth of the
potential independently from $u_{0p}$. In Fig. \ref{fig4}, we plot this potential for a number of $\gamma_p$ values.
Interestingly, for $\gamma$ values deep in the flat-top regime,
namely close to $-1$, the potential develops a two-well structure
with minima located at
$\pm\frac{1}{2}\,\sqrt{\frac{g_1}{u_0}}\,\ln\left[\frac{\sqrt{\gamma\,(8+17\gamma+9\gamma^2)}}{2(1+\gamma)}-\frac{3\gamma}{2\sqrt{1+\gamma}}-\frac{1}{\sqrt{1+\gamma}}\right]$.
The depth at the minima equals $u_0$, and the depth at the central
peak is equal to $-3u_0/3\gamma$. This double-well potential, as we
will see later, leads to reflectionless macroscopic quantum tunnelling.

In conclusion, a spectrum of solutions is obtained in terms of a
single parameter, namely $\gamma$. A spectrum of reflectionless
potentials is also obtained in terms of a single
parameter, $\gamma_p$.

\section{Resonant reflectionless scattering of flat-top and thin-top solitons}
\label{scatsec} 
\begin{figure}[H]\centering
    \includegraphics[width=5cm,clip]{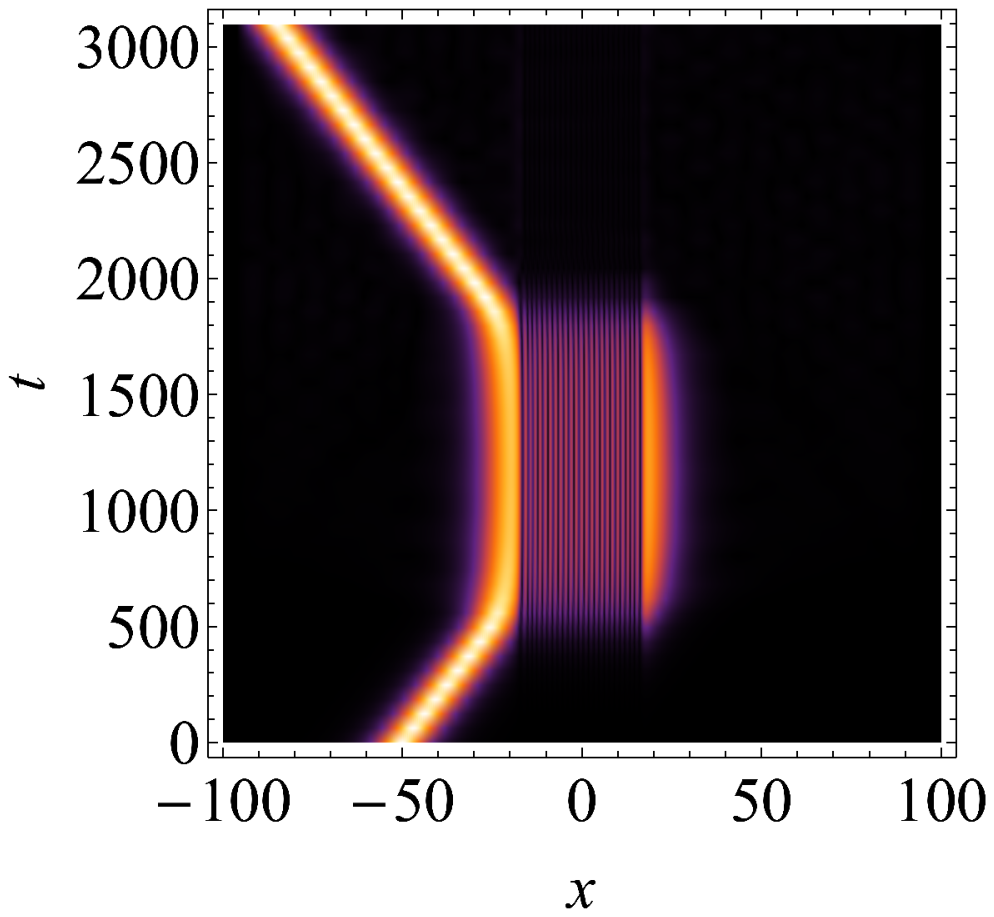}
    \begin{picture}(5,5)(5,5)
    \put(-28,119) {\color{black}{{\fcolorbox{white}{white}{\textbf{(a)}}}}}
    \end{picture}
    \includegraphics[width=5cm,clip]{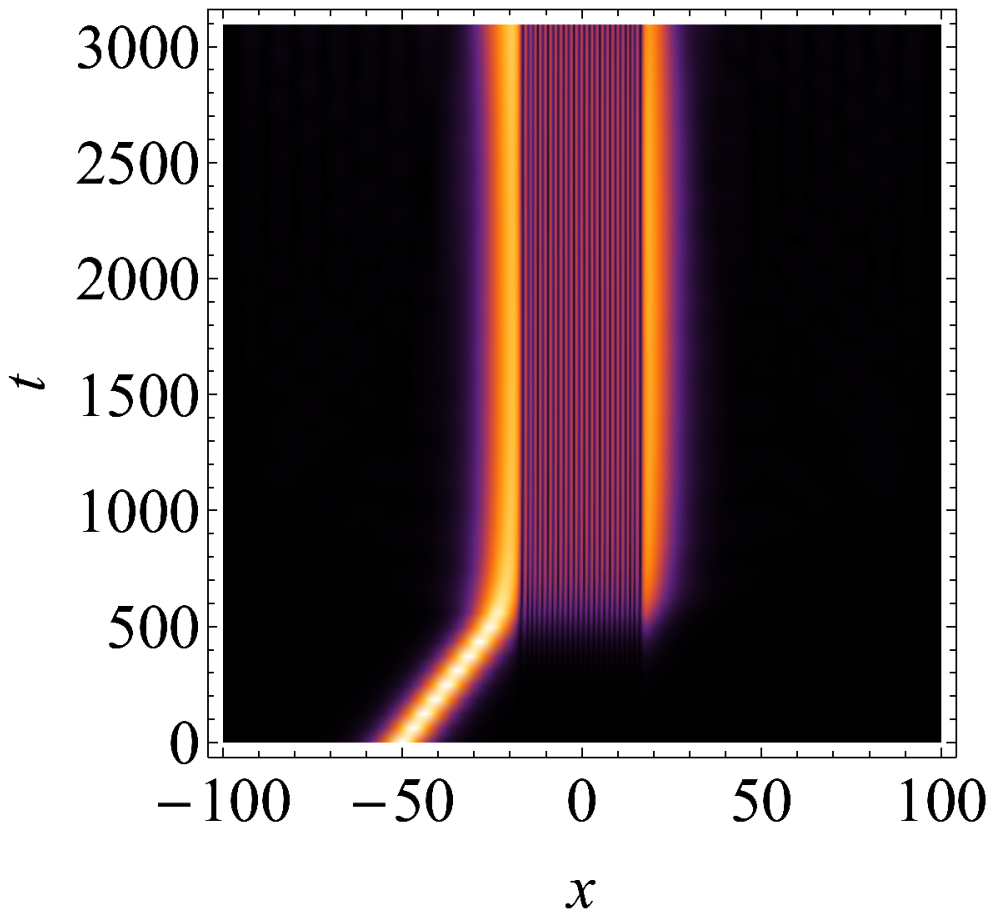}
    \begin{picture}(5,5)(5,5)
    \put(-29,119) {\color{black}{{\fcolorbox{white}{white}{\textbf{(b)}}}}}
    \end{picture}
    \includegraphics[width=5cm,clip]{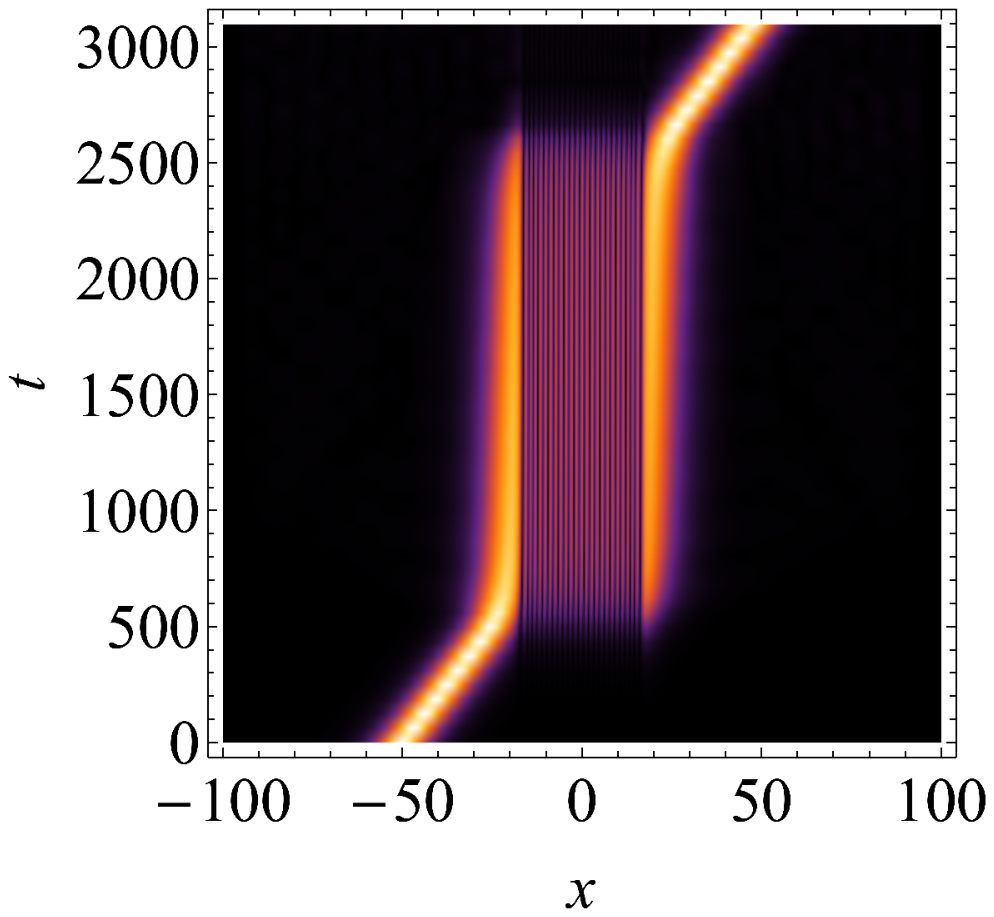}
    \begin{picture}(5,5)(5,5)
    \put(-102,119) {\color{black}{{\fcolorbox{white}{white}{\textbf{(c)}}}}}
    \end{picture}
    \includegraphics[width=9cm,clip]{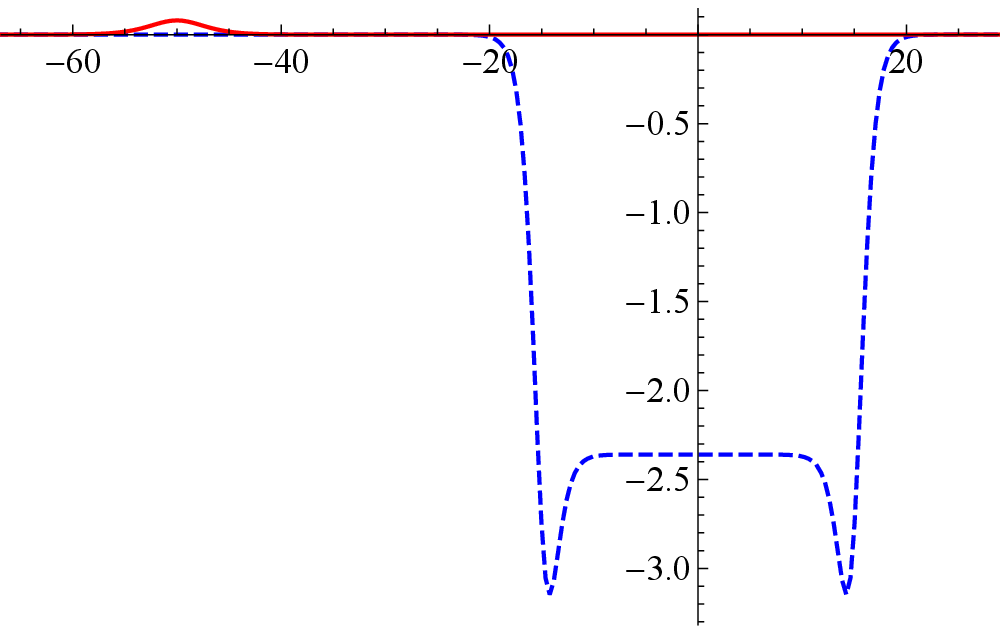}
    \begin{picture}(5,5)(5,5)
    \put(-257,123) {\color{black}{{\fcolorbox{white}{white}{\textbf{(d)}}}}}
    \put(-220,137) {\color{black}{{\fcolorbox{white}{white}{\large\textbf{$|\psi|$}}}}}
            \put(-205,158) {\color{black}{\large{$\longrightarrow$}}}
    \put(-89,-7) {\color{black}{{\fcolorbox{white}{white}{\large\textbf{$V(x)$}}}}}
    \end{picture}
    \vspace*{0.7cm}
\caption{Resonant scattering of a bright soliton (\ref{solbs}) by a
wide reflectionless potential well, given by (\ref{refpt2}), with a
varying potential well depth (a) $V_0=2.39417$, (b) $V_0=2.39418$,
and (c) $V_0=2.39419$. The number of  nodes in the formed trapped
mode is 24 nodes. (d) Solution and potential profiles.   The arrow
shows the direction of motion for the incident soliton. Other
parameters: $u_0=0.04$, $u_{0p}=0.2$, $\gamma=0$,
$\gamma_p=-0.9999999999999998$, $v_0=0.05$, $g_1=0.5$, $g_2=1.0$,
$g_{1p}=0.5$, $g_{2p}=1.0$.}
    \label{fig5}
\end{figure}

\begin{figure}[H]\centering
    \includegraphics[width=5cm,clip]{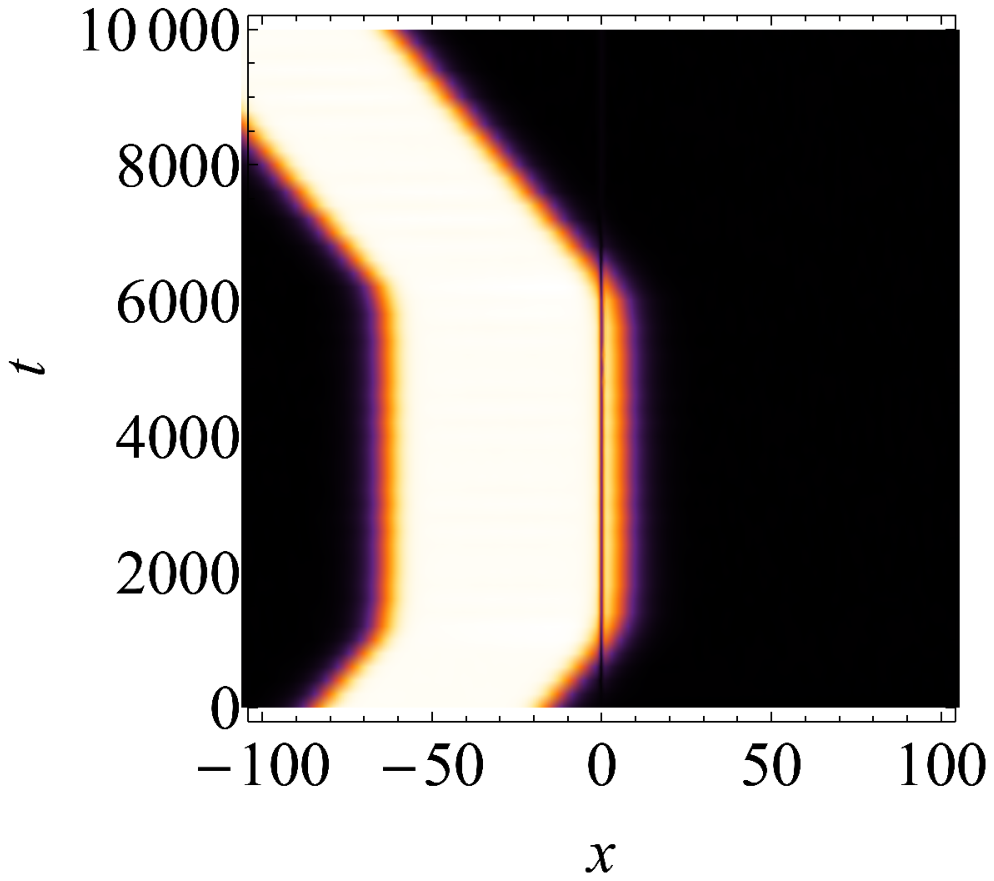}
    \begin{picture}(5,5)(5,5)
    \put(-28,114) {\color{black}{{\fcolorbox{white}{white}{\textbf{(a)}}}}}
    \end{picture}
    \includegraphics[width=5cm,clip]{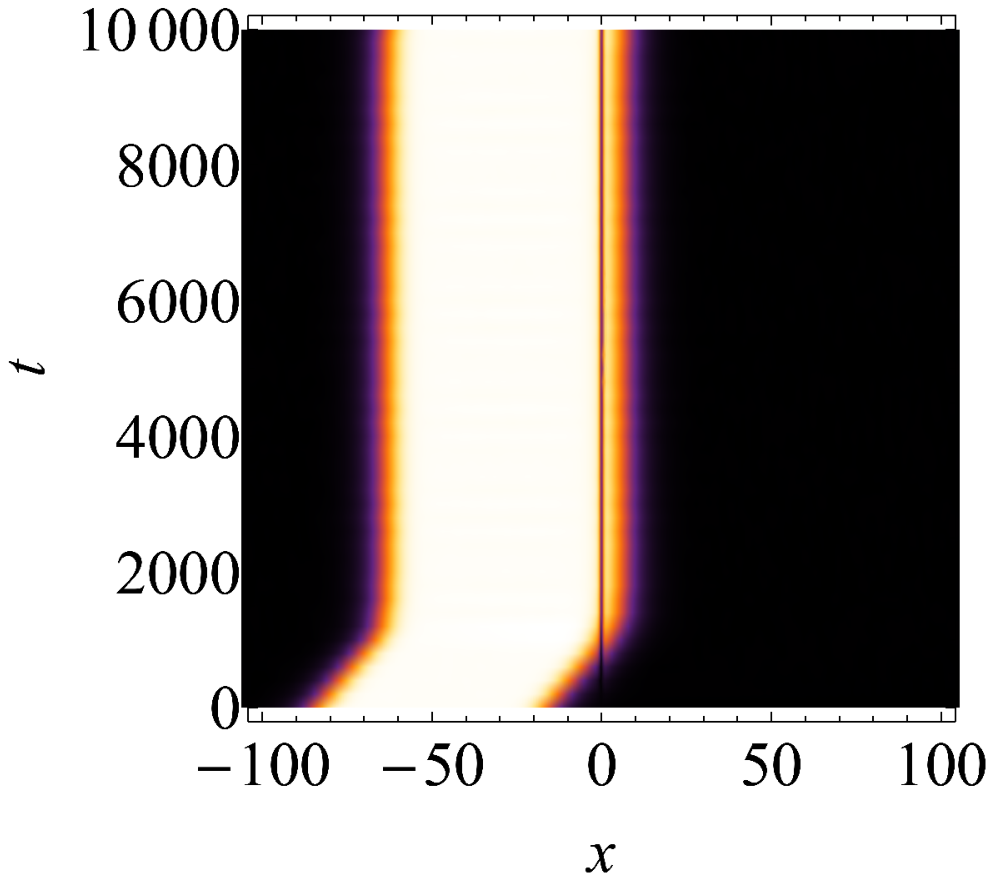}
    \begin{picture}(5,5)(5,5)
    \put(-28,114) {\color{black}{{\fcolorbox{white}{white}{\textbf{(b)}}}}}
    \end{picture}
    \includegraphics[width=5cm,clip]{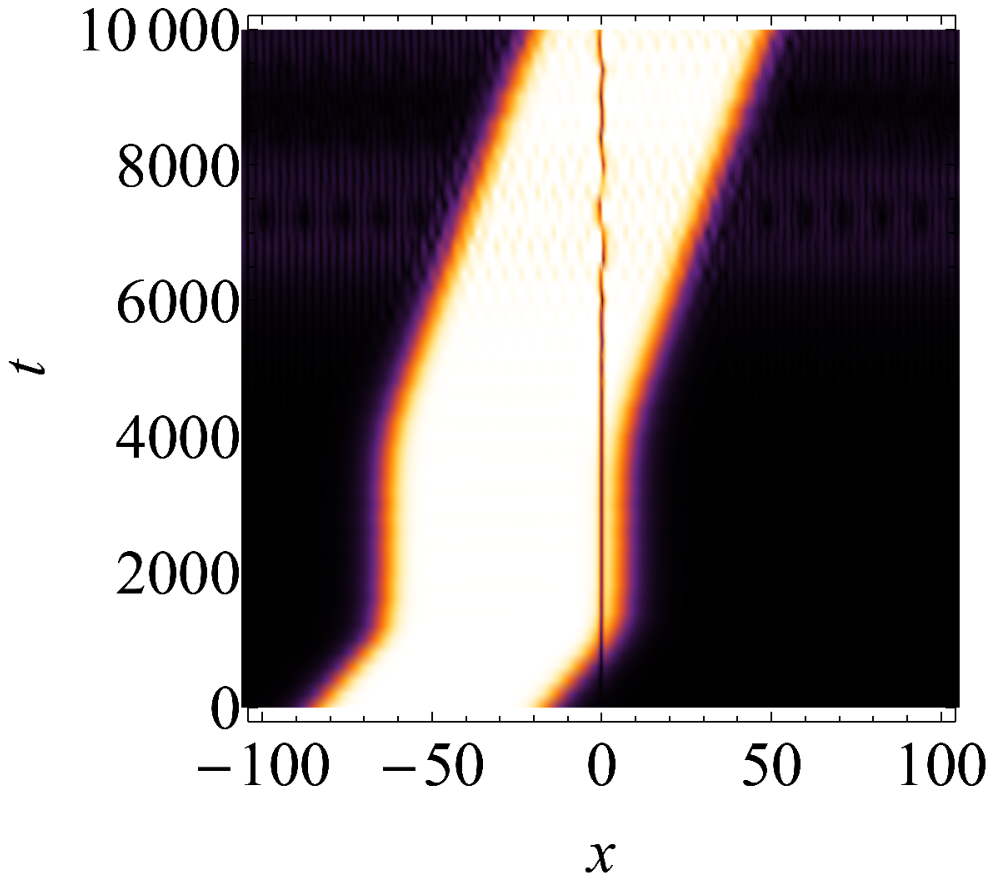}
    \begin{picture}(5,5)(5,5)
    \put(-98,114) {\color{black}{{\fcolorbox{white}{white}{\textbf{(c)}}}}}
    \end{picture}
    \includegraphics[width=9cm,clip]{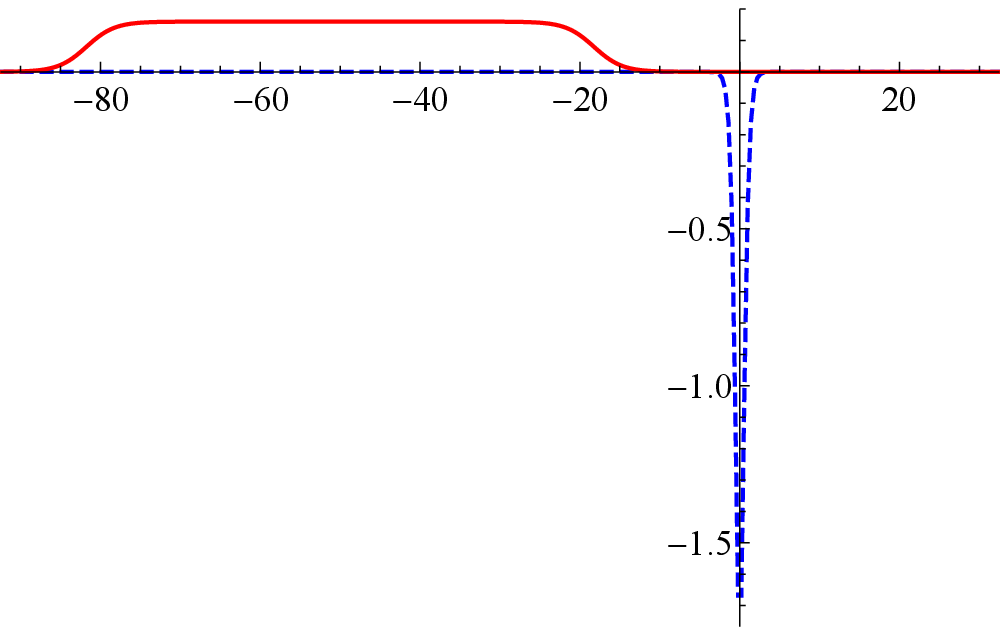}
    \begin{picture}(5,5)(5,5)
    \put(-253,120) {\color{black}{{\fcolorbox{white}{white}{\textbf{(d)}}}}}
    \put(-179,127) {\color{black}{{\fcolorbox{white}{white}{\large\textbf{$|\psi|$}}}}}
            \put(-107,154) {\color{black}{\large{$\longrightarrow$}}}
    \put(-85,-7) {\color{black}{{\fcolorbox{white}{white}{\large\textbf{$V(x)$}}}}}
    \end{picture}
    \vspace*{0.7cm}
    \caption{Resonant scattering of a flat-top soliton (\ref{solft}) by a P\"oschl-Teller potential well (\ref{ptpot}) with a varying potential well depth (a) $V_0=0.895586$, (b) $V_0=0.8955860015377$, and (c) $V_0=0.895587$. (d) The solution and potential profiles.  A single-node trapped mode is formed. The arrow shows the direction of motion for the incident soliton. Other parameters: $u_0=0.04$, $u_{0p}=2.44779$, $\gamma=-0.999999999999999$, $\gamma_p=0$, $v_0=0.017$, $g_1=0.5$, $g_2=1.0$, $g_{1p}=0.5$, $g_{2p}=1.0$.}
    \label{fig6}
\end{figure}

\begin{figure}[H]\centering
    \includegraphics[width=5cm,clip]{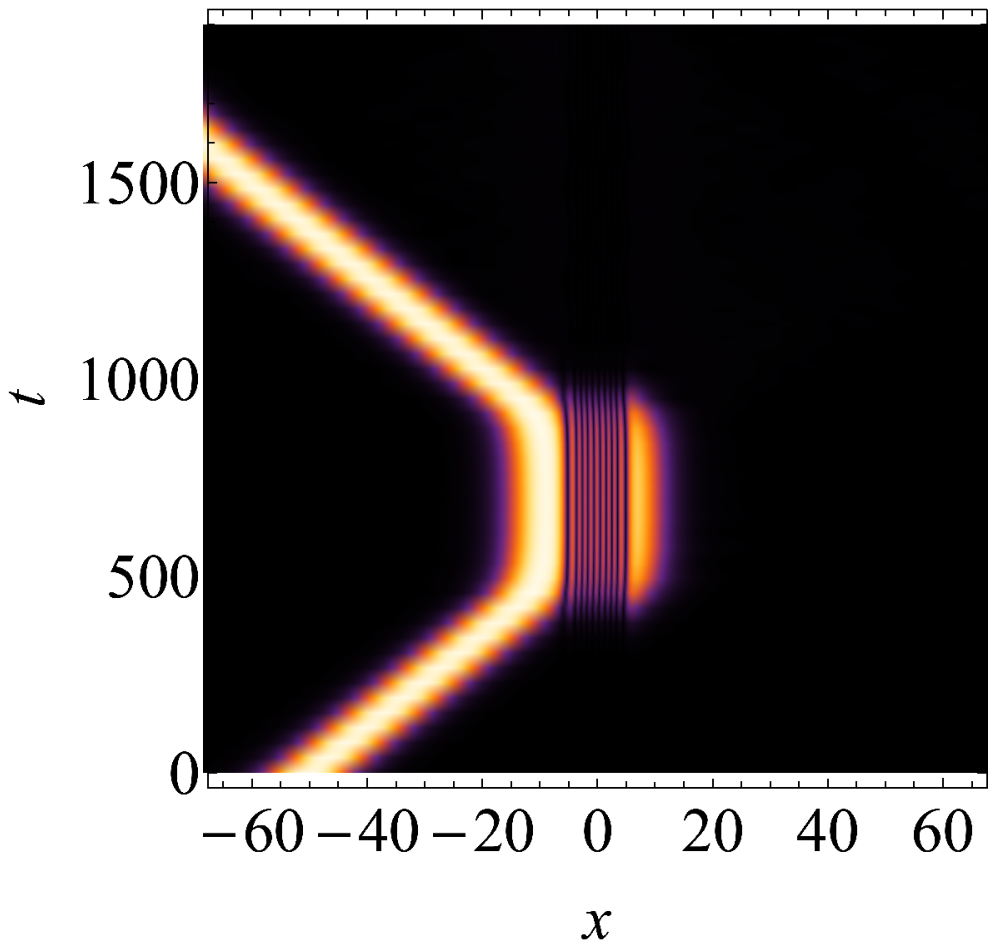}
    \begin{picture}(5,5)(5,5)
    \put(-24,122) {\color{black}{{\fcolorbox{white}{white}{\textbf{(a)}}}}}
    \end{picture}
    \includegraphics[width=5cm,clip]{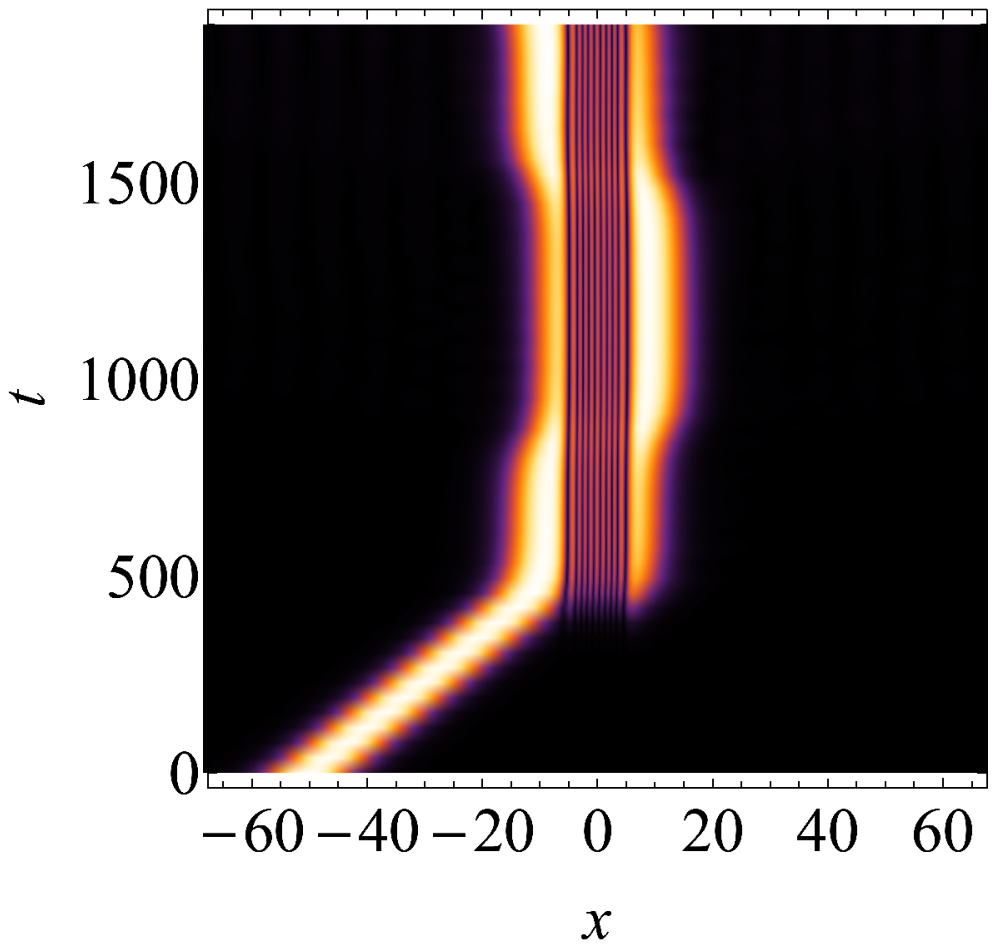}
    \begin{picture}(5,5)(5,5)
    \put(-24,122) {\color{black}{{\fcolorbox{white}{white}{\textbf{(b)}}}}}
    \end{picture}
    \includegraphics[width=5cm,clip]{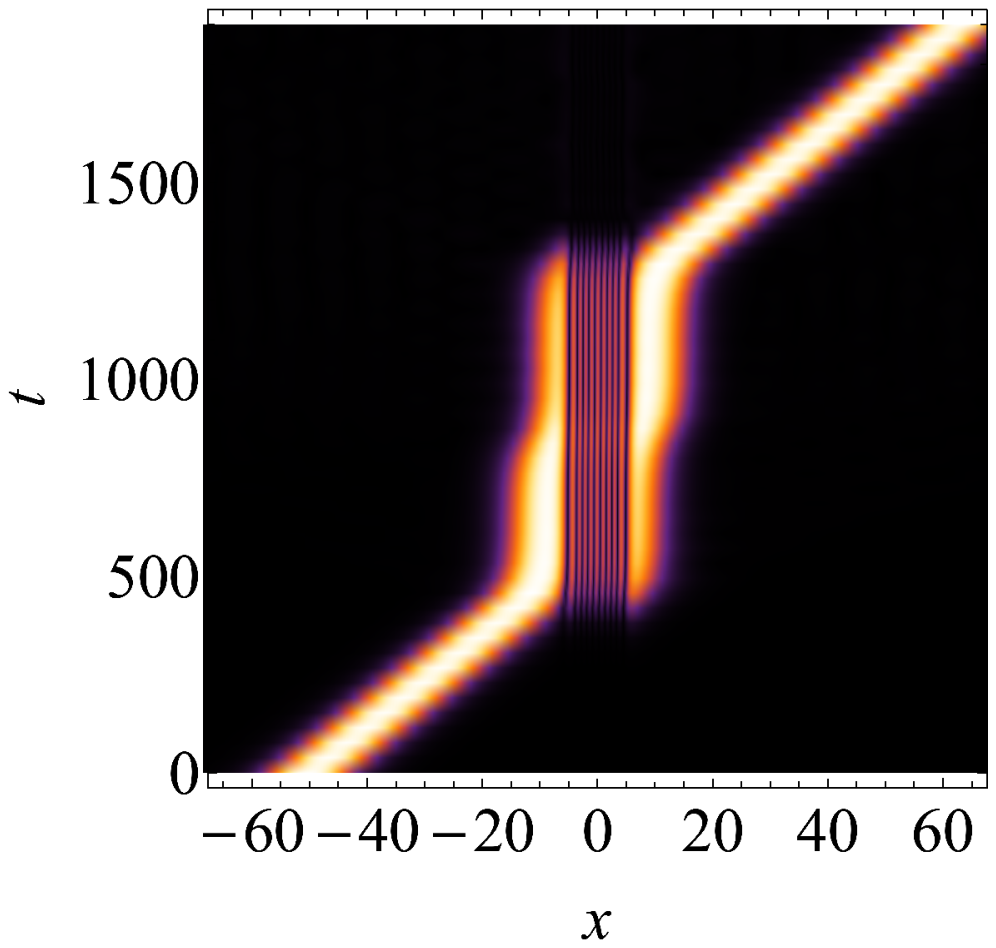}
    \begin{picture}(5,5)(5,5)
    \put(-104,122) {\color{black}{{\fcolorbox{white}{white}{\textbf{(c)}}}}}
    \end{picture}
    \includegraphics[width=9cm,clip]{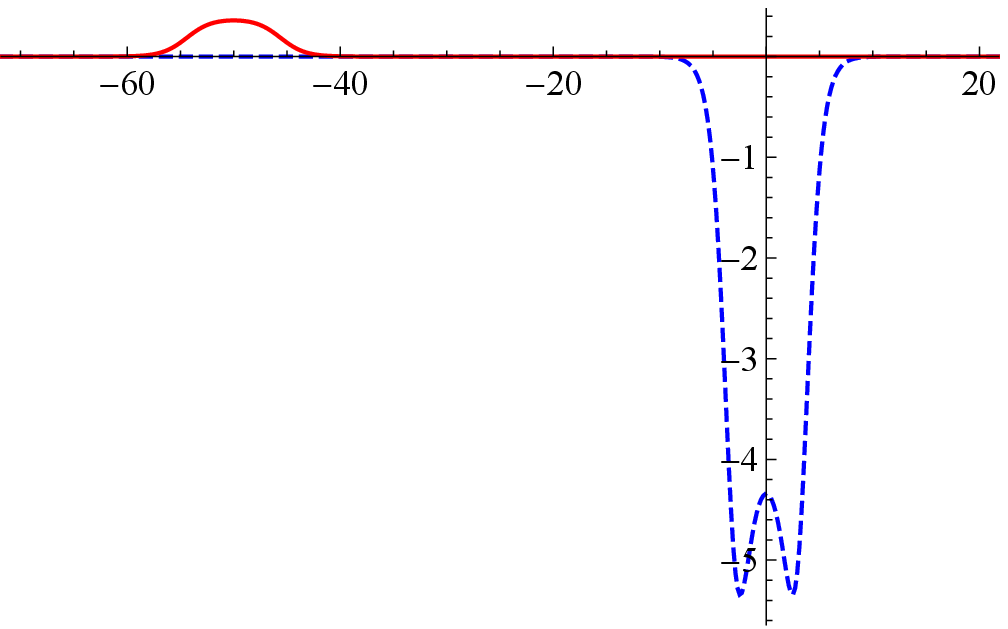}
    \begin{picture}(5,5)(5,5)
    \put(-260,120) {\color{black}{{\fcolorbox{white}{white}{\textbf{(d)}}}}}
    \put(-204,129) {\color{black}{{\fcolorbox{white}{white}{\large\textbf{$|\psi|$}}}}}
        \put(-188,156) {\color{black}{\large{$\longrightarrow$}}}
    \put(-78,-9) {\color{black}{\large\textbf{$V(x)$}}}
    \end{picture}
    \vspace*{0.7cm}
    \caption{Resonant scattering of a flat-top soliton (\ref{solft}) by a wide reflectionless potential well with a varying potential well depth (a) $V_0=4.0$, (b) $V_0=4.00002762$, and (c) $V_0=4.000029$. (d) The solution and potential profiles. The number of  nodes in the formed trapped mode is 10 nodes.  The arrow shows the direction of motion for the incident soliton. Other parameters: $u_0=0.093748$, $u_{0p}=0.2$, $\gamma=\gamma_p=-0.998$, $v_0=0.1$, $g_1=0.5$, $g_2=1.0$, $g_{1p}=0.5$, $g_{2p}=1.0$.}
    \label{fig7}
    \end{figure}

\begin{figure}[H]\centering
    \includegraphics[width=5cm,clip]{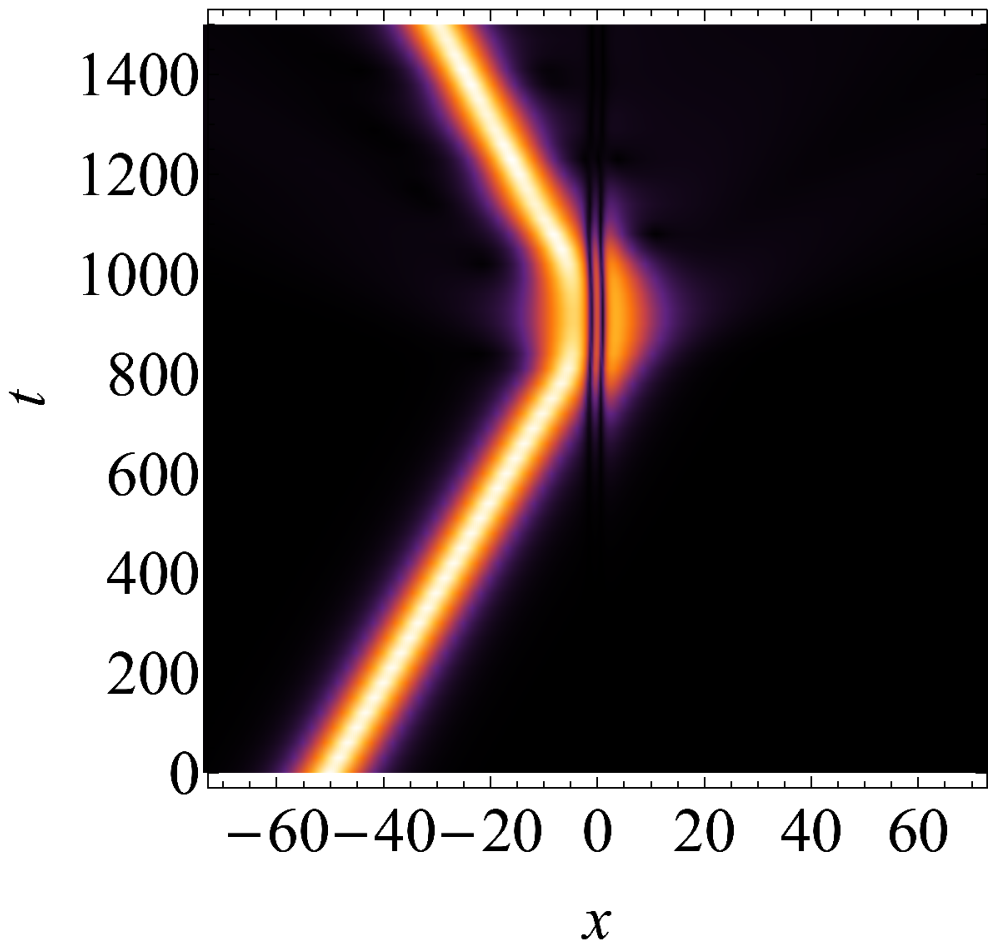}
    \begin{picture}(5,5)(5,5)
    \put(-24,122) {\color{black}{{\fcolorbox{white}{white}{\textbf{(a)}}}}}
    \end{picture}
    \includegraphics[width=5cm,clip]{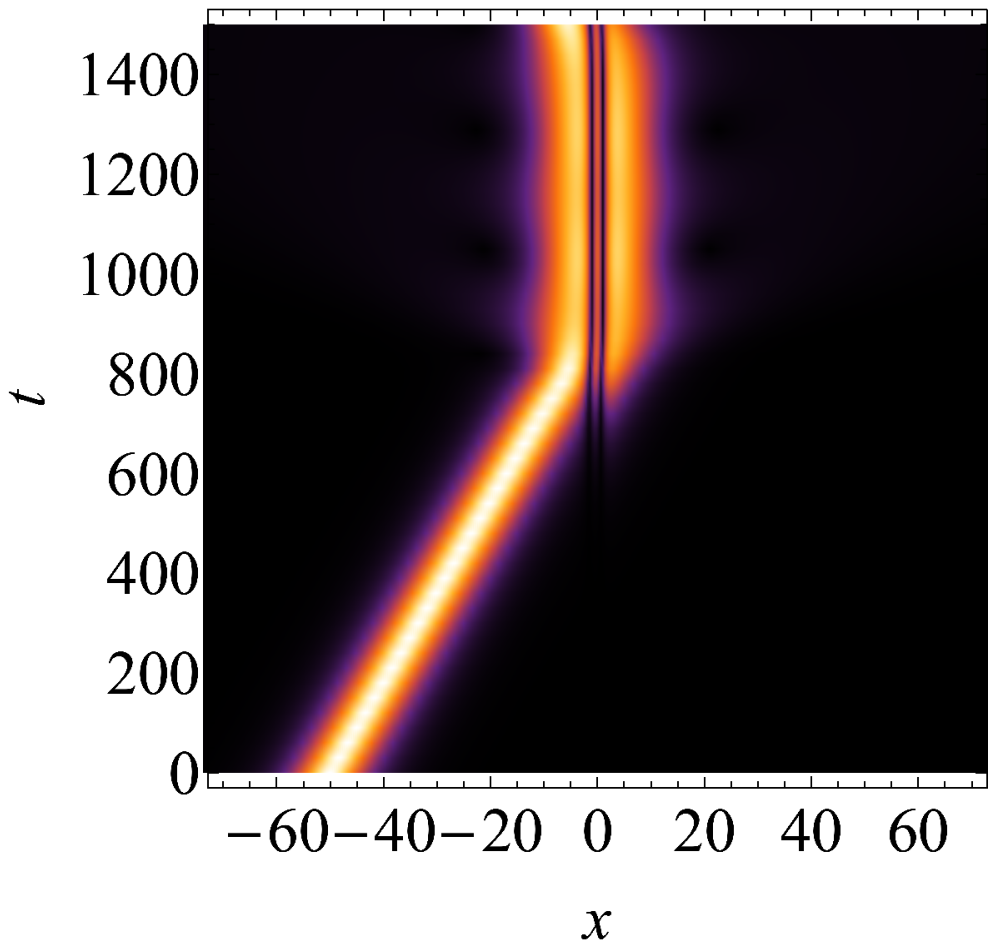}
    \begin{picture}(5,5)(5,5)
    \put(-24,122) {\color{black}{{\fcolorbox{white}{white}{\textbf{(b)}}}}}
    \end{picture}
    \includegraphics[width=5cm,clip]{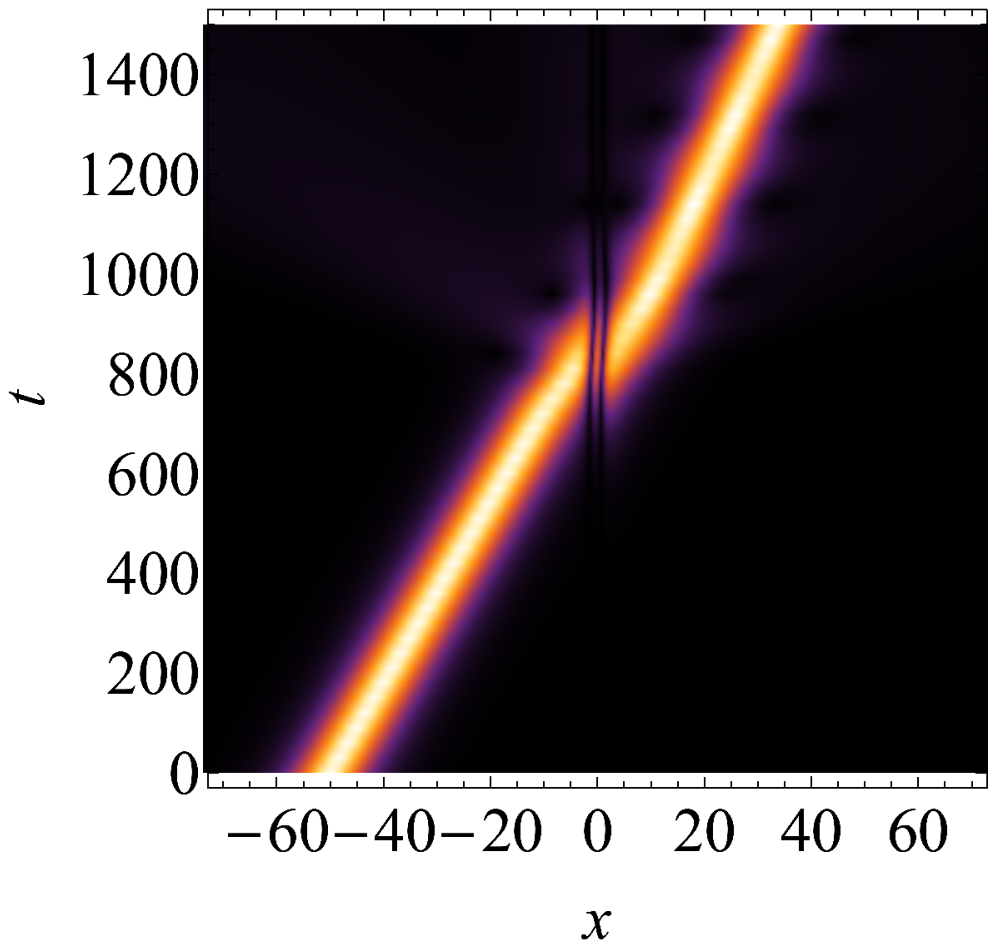}
    \begin{picture}(5,5)(5,5)
    \put(-104,122) {\color{black}{{\fcolorbox{white}{white}{\textbf{(c)}}}}}
    \end{picture}
    \includegraphics[width=9cm,clip]{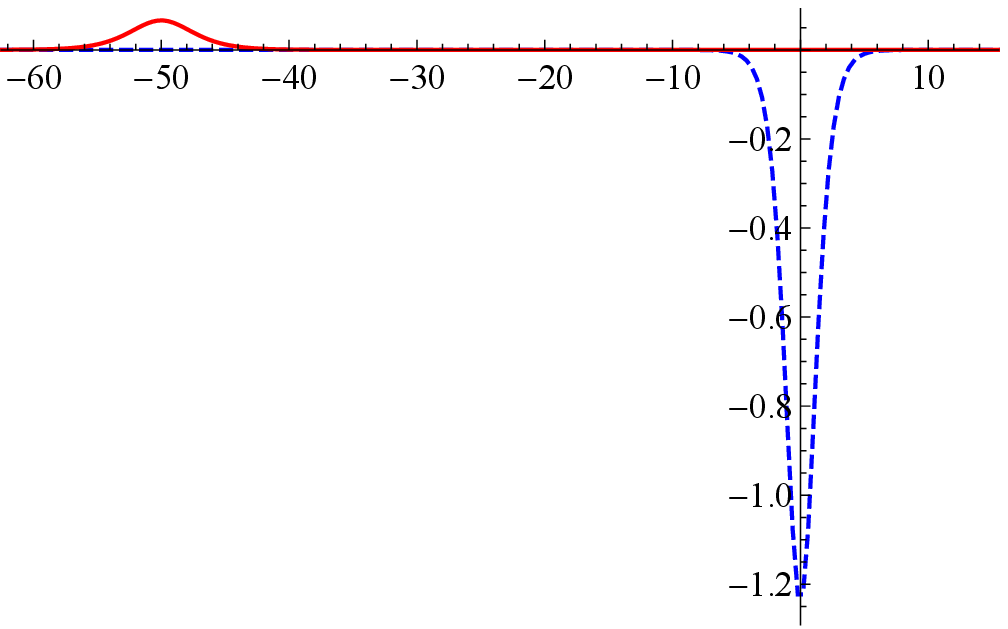}
    \begin{picture}(5,5)(5,5)
    \put(-260,120) {\color{black}{{\fcolorbox{white}{white}{\textbf{(d)}}}}}
    \put(-222,129) {\color{black}{{\fcolorbox{white}{white}{\large\textbf{$|\psi|$}}}}}
    \put(-208,156) {\color{black}{\large{$\longrightarrow$}}}
    \put(-66,-9) {\color{black}{\large\textbf{$V(x)$}}}
    \end{picture}
    \vspace*{0.7cm}
    \caption{Resonant scattering of a thin-top soliton (\ref{soltt}) by a P\"oschl-Teller potential well with a varying potential well depth (a) $V_0=0.62$, (b) $V_0=0.621357995$, and (c) $V_0=0.635$. (d) The solution and potential profiles.  A double-node trapped mode is formed.  The arrow shows the direction of motion for the incident soliton. Other parameters: $u_0=0.04$, $u_{0p}=0.2$, $\gamma=1.0$, $\gamma_p=0$, $v_0=0.055$, $g_1=0.5$, $g_2=1.0$, $g_{1p}=0.5$, $g_{2p}=1.0$.}
    \label{fig10}
\end{figure}

\begin{figure}[H]\centering
    \includegraphics[width=5cm,clip]{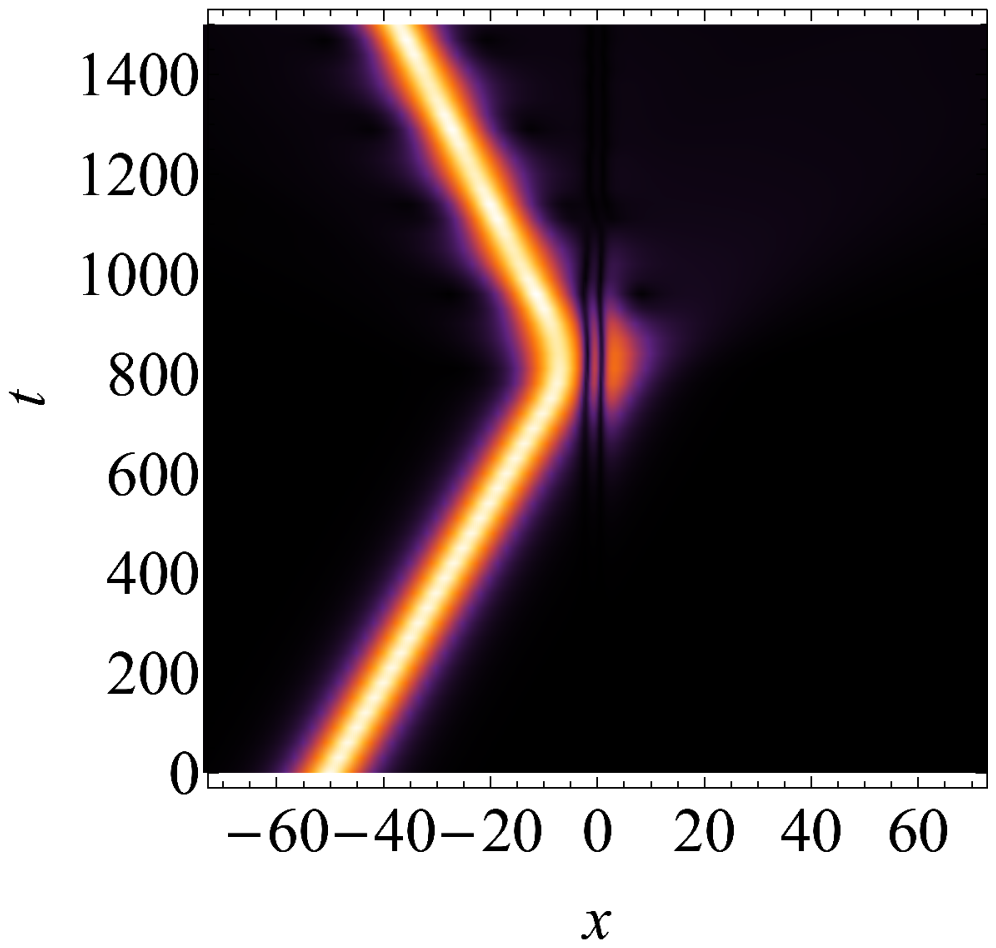}
    \begin{picture}(5,5)(5,5)
    \put(-24,122) {\color{black}{{\fcolorbox{white}{white}{\textbf{(a)}}}}}
    \end{picture}
    \includegraphics[width=5cm,clip]{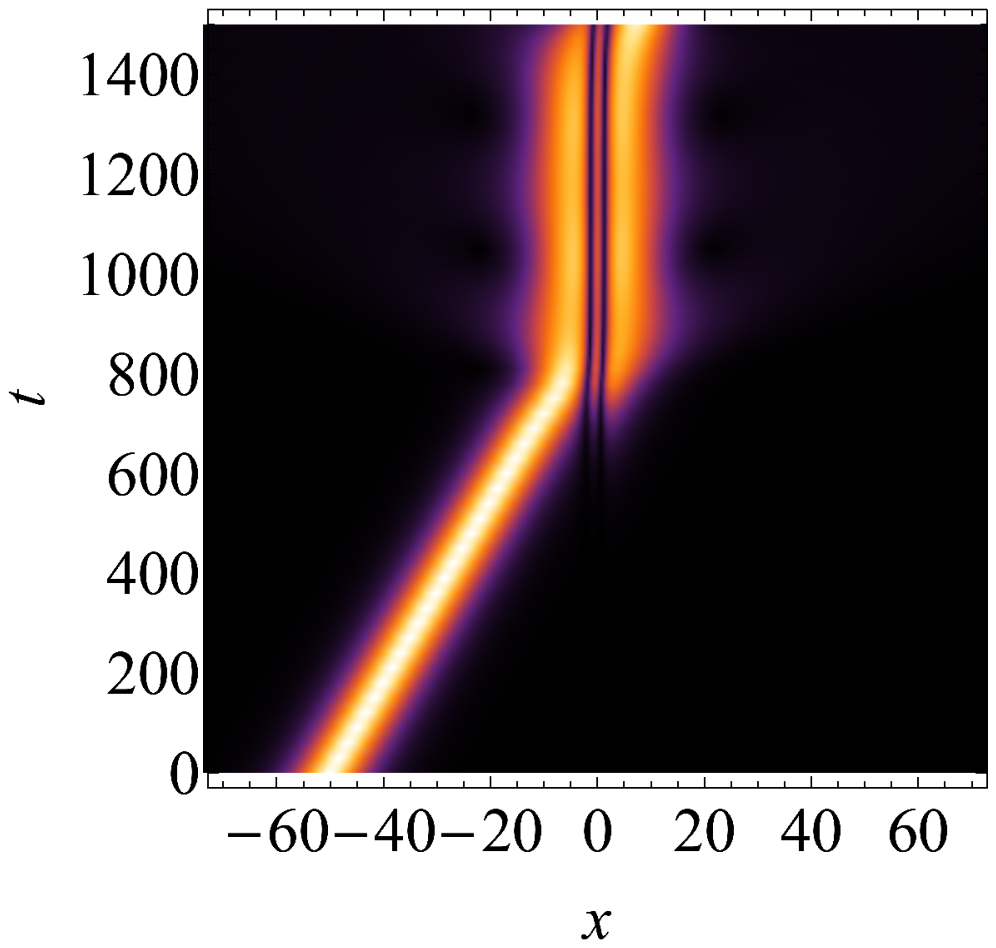}
    \begin{picture}(5,5)(5,5)
    \put(-24,122) {\color{black}{{\fcolorbox{white}{white}{\textbf{(b)}}}}}
    \end{picture}
    \includegraphics[width=5cm,clip]{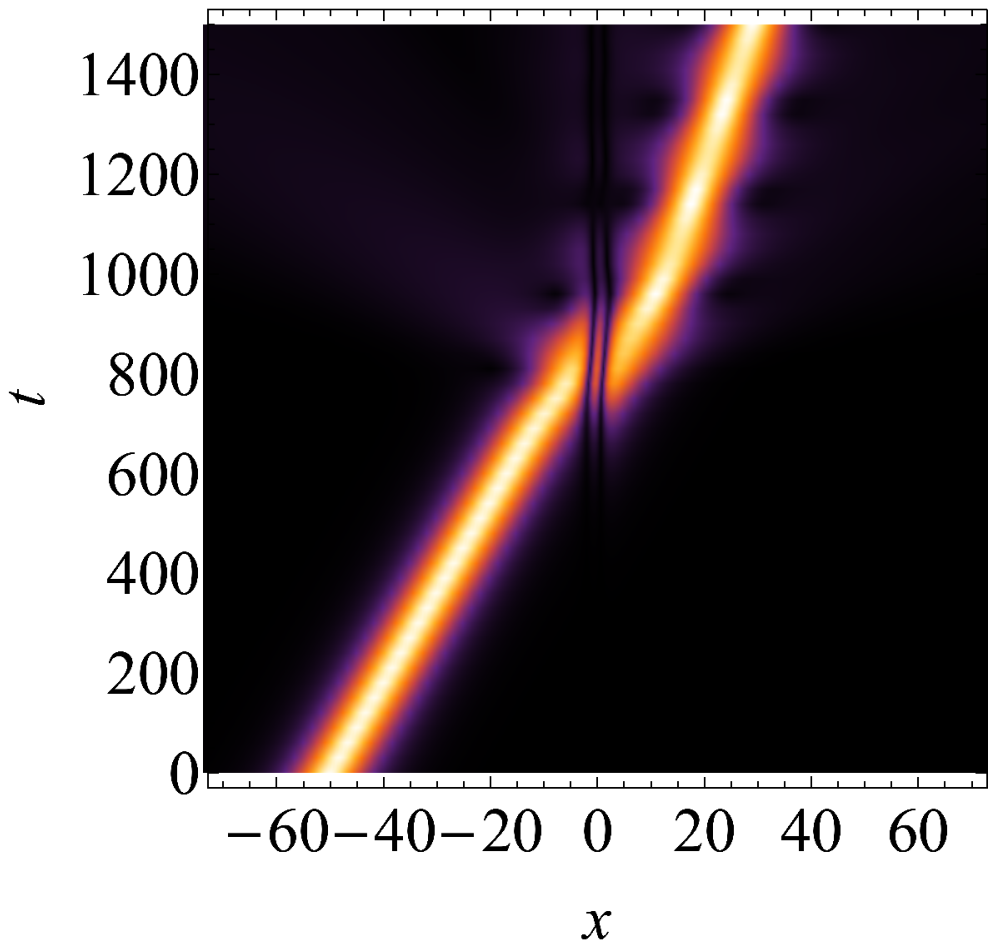}
    \begin{picture}(5,5)(5,5)
    \put(-104,122) {\color{black}{{\fcolorbox{white}{white}{\textbf{(c)}}}}}
    \end{picture}
    \includegraphics[width=9cm,clip]{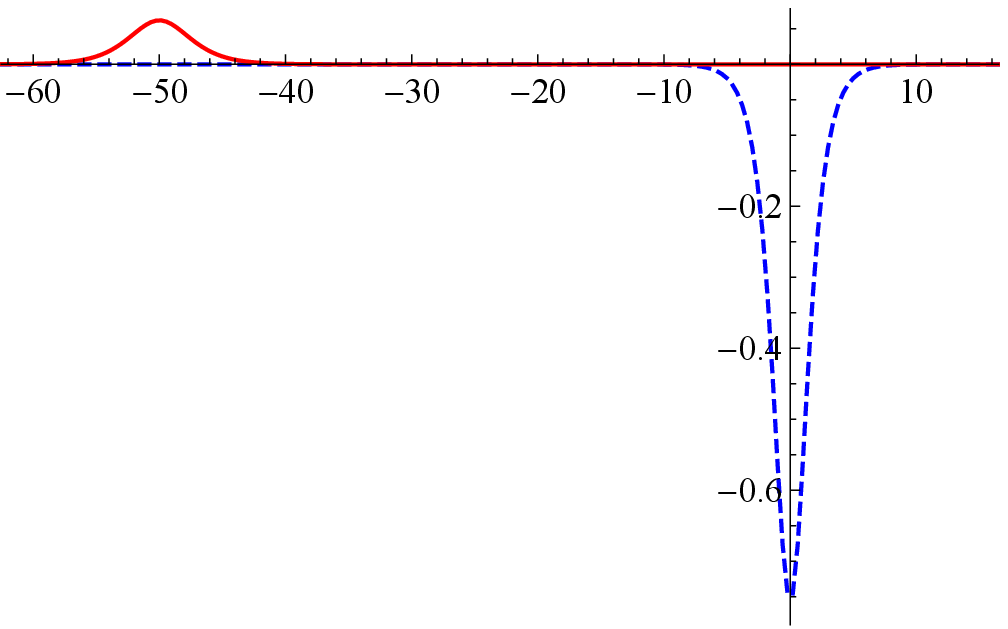}
    \begin{picture}(5,5)(5,5)
    \put(-260,120) {\color{black}{{\fcolorbox{white}{white}{\textbf{(d)}}}}}
    \put(-223,126) {\color{black}{{\fcolorbox{white}{white}{\large\textbf{$|\psi|$}}}}}
    \put(-208,156) {\color{black}{\large{$\longrightarrow$}}}
    \put(-66,-9) {\color{black}{\large\textbf{$V(x)$}}}
    \end{picture}
    \vspace*{0.7cm}
    \caption{Resonant scattering of a thin-top soliton (\ref{soltt}) by a thin reflectionless  potential well with a varying potential well depth (a) $V_0=0.38$, (b) $V_0=0.3841987$, and (c) $V_0=0.395$. (d) The solution and potential profiles.  A double-node trapped mode is formed.  The arrow shows the direction of motion for the incident soliton. Other parameters: $u_0=0.04$, $u_{0p}=0.1$, $\gamma=\gamma_p=1.5$,  $v_0=0.055$, $g_1=0.5$, $g_2=1.0$, $g_{1p}=0.5$, $g_{2p}=1.0$.}
    \label{fig11}
\end{figure}

\begin{figure}[H]\centering
    \includegraphics[width=5cm,clip]{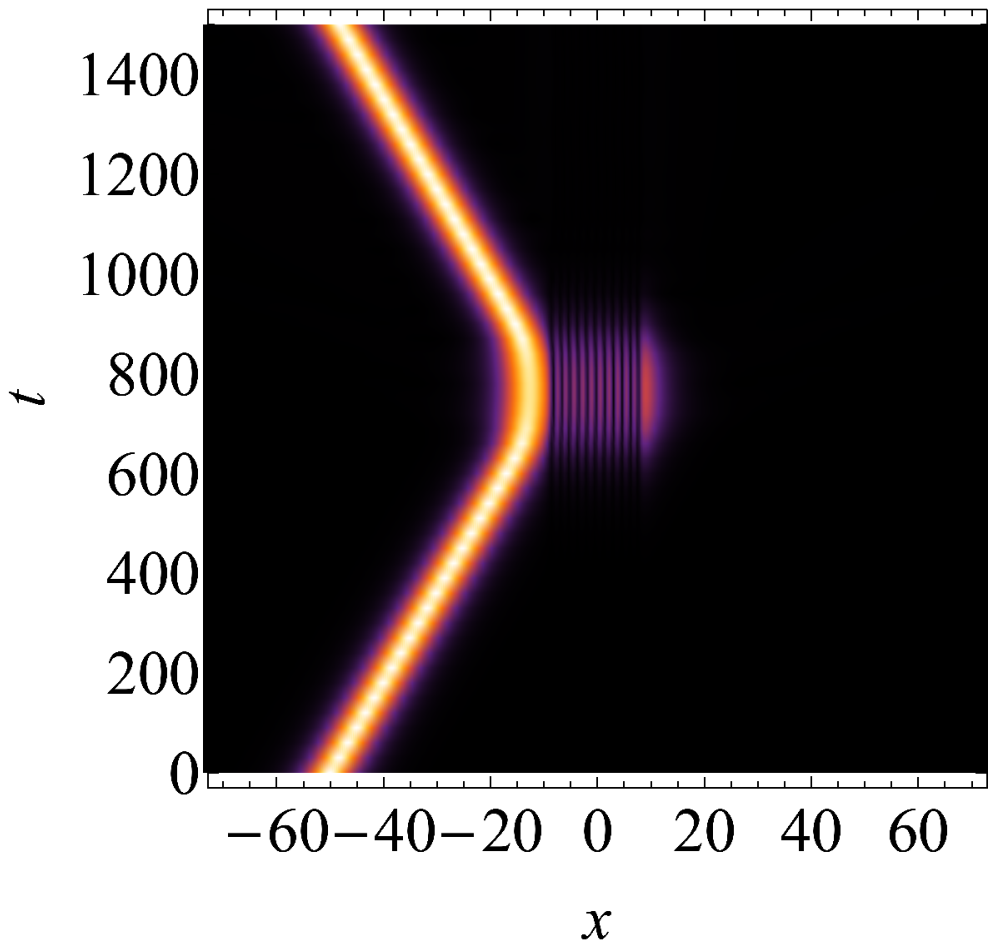}
    \begin{picture}(5,5)(5,5)
    \put(-24,122) {\color{black}{{\fcolorbox{white}{white}{\textbf{(a)}}}}}
    \end{picture}
    \includegraphics[width=5cm,clip]{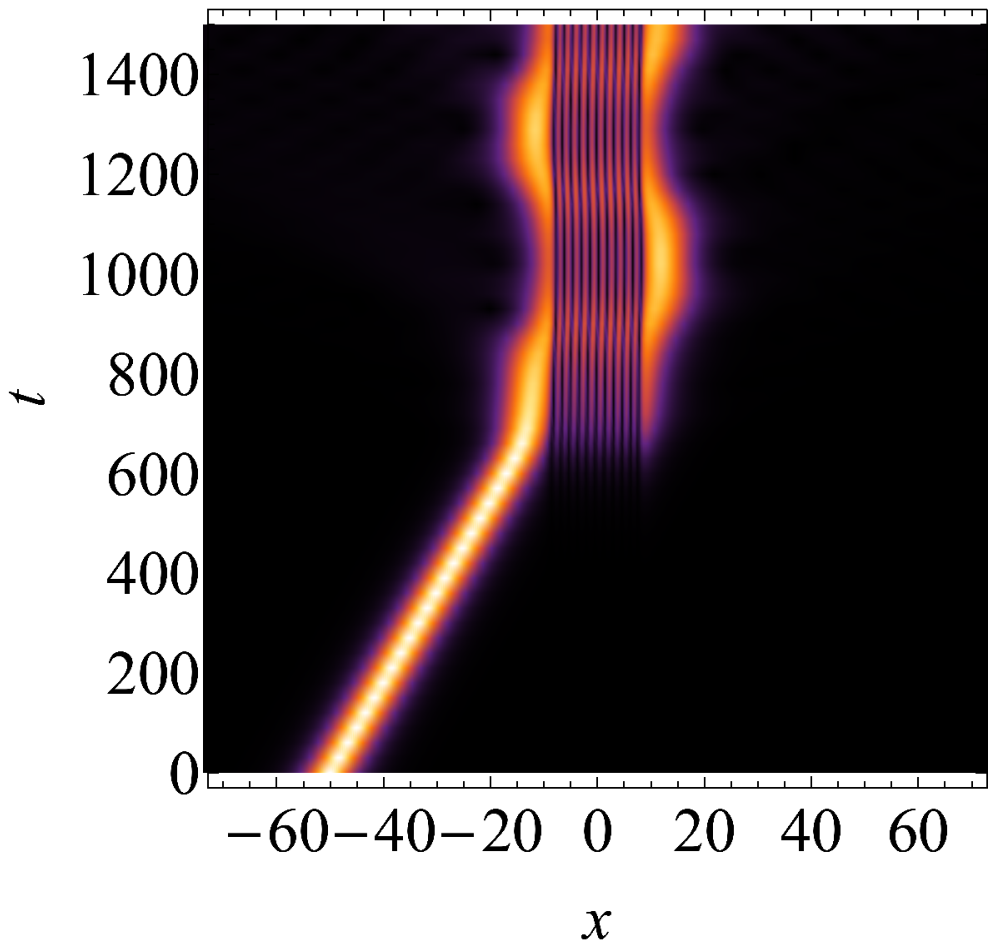}
    \begin{picture}(5,5)(5,5)
    \put(-24,122) {\color{black}{{\fcolorbox{white}{white}{\textbf{(b)}}}}}
    \end{picture}
    \includegraphics[width=5cm,clip]{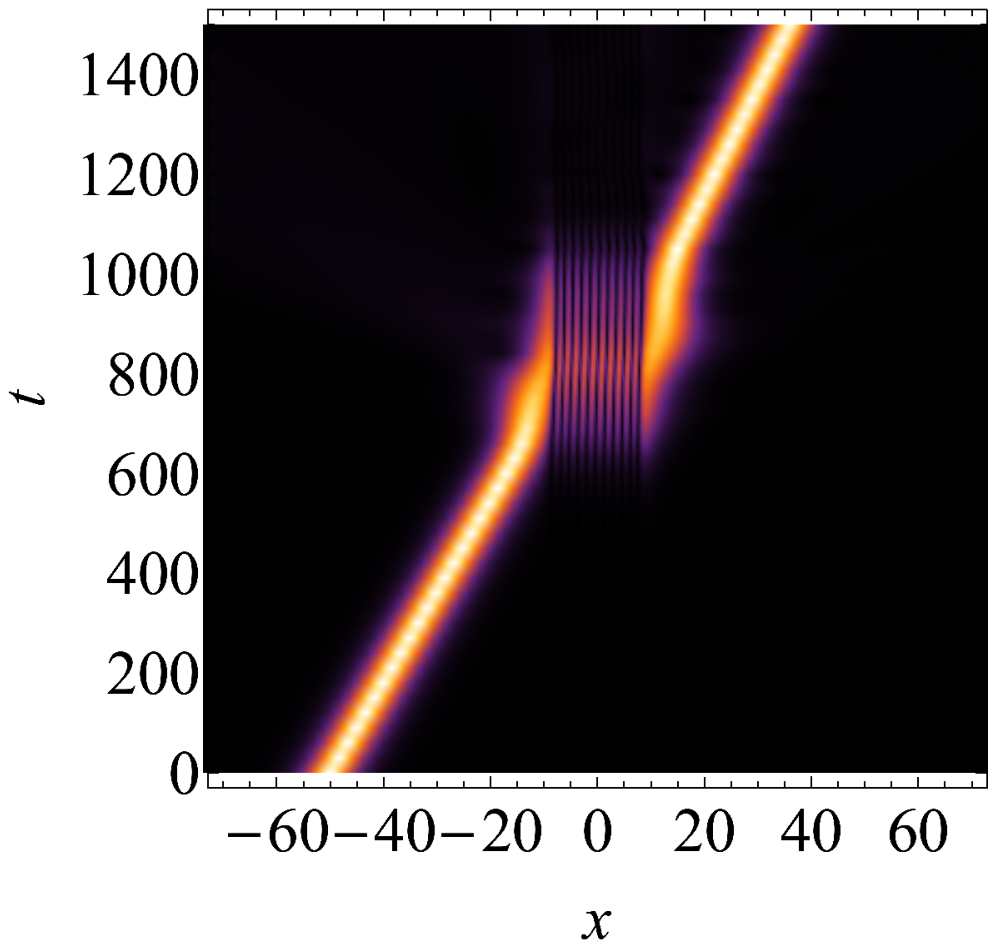}
    \begin{picture}(5,5)(5,5)
    \put(-104,122) {\color{black}{{\fcolorbox{white}{white}{\textbf{(c)}}}}}
    \end{picture}
    \includegraphics[width=9cm,clip]{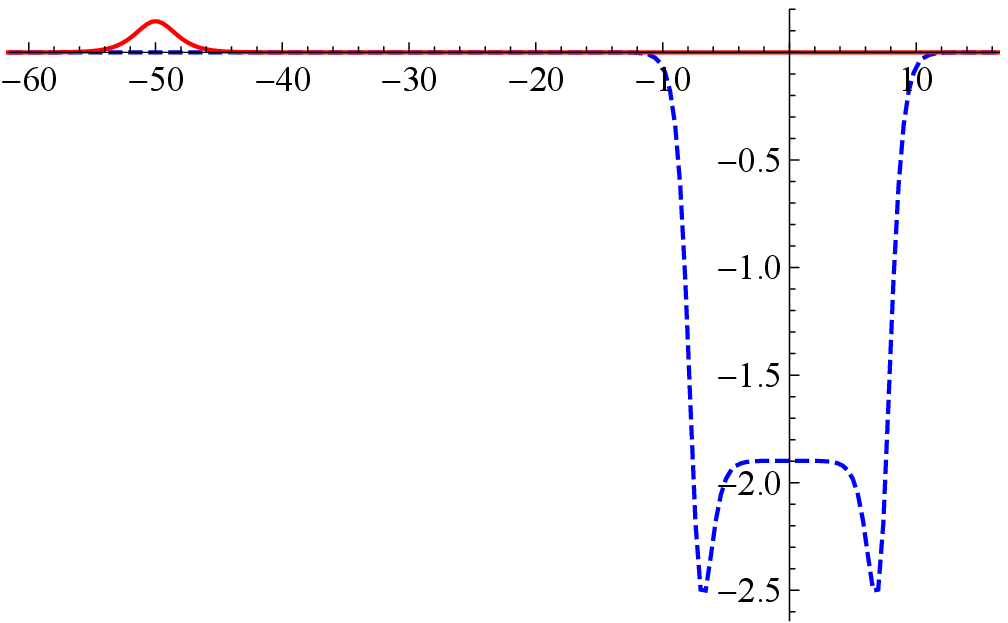}
    \begin{picture}(5,5)(5,5)
    \put(-260,120) {\color{black}{{\fcolorbox{white}{white}{\textbf{(d)}}}}}
    \put(-223,126) {\color{black}{{\fcolorbox{white}{white}{\large\textbf{$|\psi|$}}}}}
    \put(-212,155) {\color{black}{\large{$\longrightarrow$}}}
    \put(-66,-9) {\color{black}{\large\textbf{$V(x)$}}}
    \end{picture}
    \vspace*{0.7cm}
    \caption{Resonant scattering of a thin-top soliton (\ref{soltt}) by a wide reflectionless
    potential well with a varying potential well depth (a) $V_0=1.897$, (b) $V_0=1.898$, and (c) $V_0=1.9$. (d) The solution and potential profiles.  The number of  nodes in the formed trapped mode is 11 nodes.  The arrow shows the direction of motion for the incident soliton. Other parameters: $u_0=0.0937480$, $u_{0p}=0.5$, $\gamma=1.5$, $\gamma_p=-0.9999999999$,  $v_0=0.055$, $g_1=0.5$, $g_2=1.0$, $g_{1p}=0.7$, $g_{2p}=1.0$.}
    \label{fig12}
\end{figure}

The purpose of this section is to confirm the
reflectionless property of the potential derived in the previous
section. This is performed through scattering of flat-top and
thin-top solitons by the potential. In view of the fact that a
spectrum of solutions and a corresponding spectrum of reflectionless
potentials exist, we consider here  possible combinations of
solutions and potentials, as follows:\\\\
{\bf Bright soliton scattered by wide reflectionless potential well, $\gamma=0$, $\gamma_p=2\times10^{-16}-1$}:\\
This case shows an interesting resonant scattering of a bright
soliton by a wide reflectionless potential where a multi-node
trapped mode is formed. The sharp transition between full quantum
reflection and full transmission is evident in Fig. \ref{fig5}.
It should be noted
here that a square potential well with a similar width and depth would
generate a considerable amount of radiation, which indicates the
unique reflectionless feature of the potential at hand.\\\\
{\bf Flat-top soliton scattered by P\"oschl-Teller potential well,
$\gamma=-0.999999999999999$, $\gamma_p=0$}:\\
It is interesting to see that such a huge soliton scatters
coherently by the reflectionless potential. Quantum reflection and
resonant scattering are clearly seen in Fig. \ref{fig6}.
It is known that a P\"oschl-Teller potential with large modified
width supports multi-node trapped modes, but loses its
reflectionless property. In the present case, we report a situation
where a potential with such a large width
 still results in a reflectionless scattering. \\\\
{\bf
Flat-top soliton scattered by wide reflectionless potential well, {\it macroscopic quantum tunnelling},
$\gamma=\gamma_p=-0.998$}:\\
This is a typical general case where a wide flat-top soliton is
scattered by a wide reflectionless potential. Figure \ref{fig7} shows 
reflectionless scattering together with multinode trapped modes
excitation. It is interesting to see that with such a double-potential well, macroscopic quantum tunnelling occurs through the barrier between the two wells.  This is similar to same phenomenon in Bose-Einstein condensates \cite{smerzi}.
{\bf Thin-top scattering:} In a similar manner as for the flat-top
solitons, we performed a series of simulations of scattering of
thin-top solitons by a P\"oschl-Teller potential well, shown in Fig.
\ref{fig10}, a thin reflectionless potential well shown in Fig.
\ref{fig11}, and a wide reflectionless potential well shown in Fig.
\ref{fig12}. Similar results are obtained as for  flat-top
solitons.

\begin{figure}[H]\centering
    \includegraphics[width=8.5cm,clip]{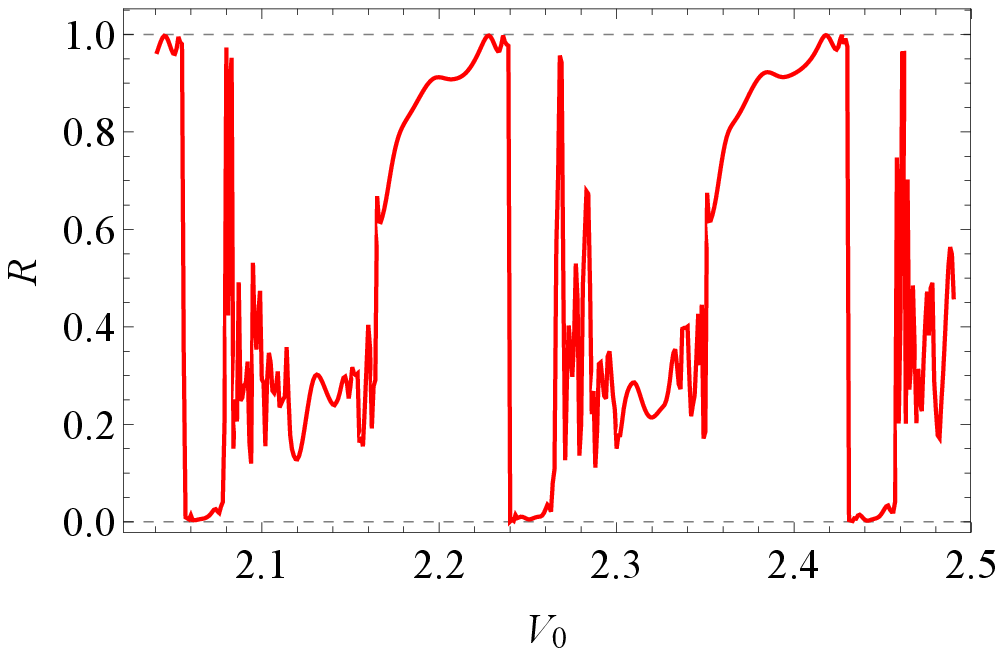}
    \begin{picture}(5,5)(5,5)
    \put(-28,132) {\color{black}{{\fcolorbox{white}{white}{\textbf{(a)}}}}}
    \end{picture}

    \includegraphics[width=8.5cm,clip]{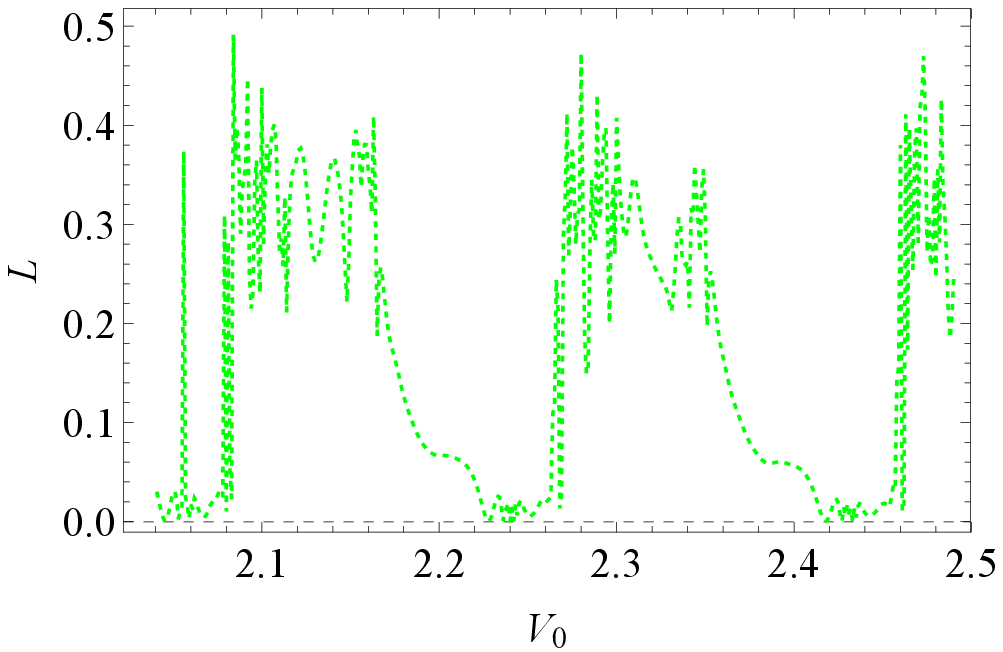}
    \begin{picture}(5,5)(5,5)
    \put(-28,132) {\color{black}{{\fcolorbox{white}{white}{\textbf{(b)}}}}}
    \end{picture}

    \includegraphics[width=8.5cm,clip]{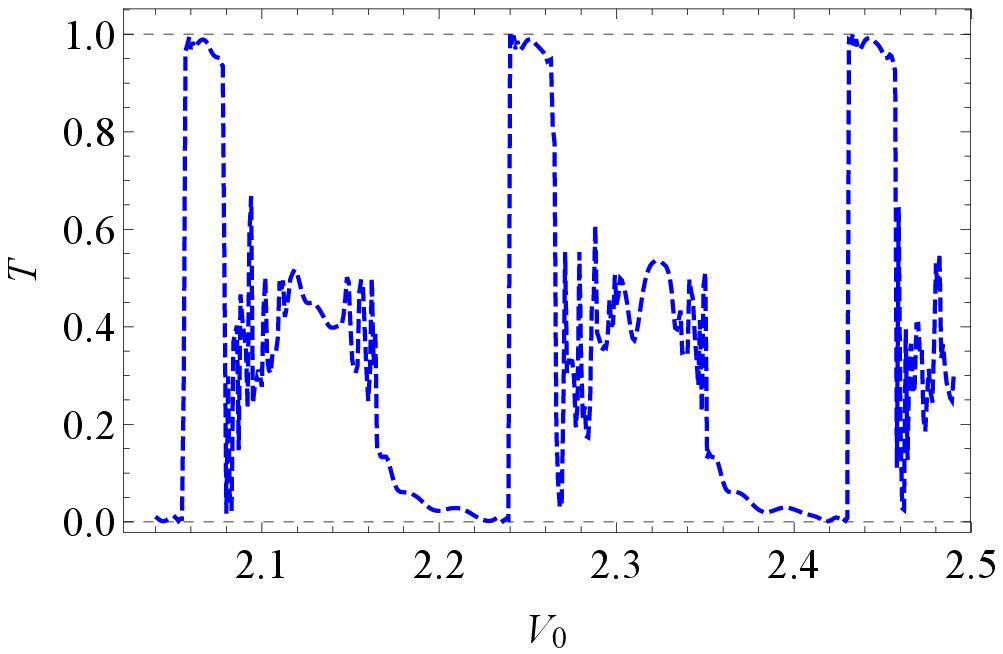}
    \begin{picture}(5,5)(5,5)
    \put(-28,132) {\color{black}{{\fcolorbox{white}{white}{\textbf{(c)}}}}}
    \end{picture}

    \includegraphics[width=8.5cm,clip]{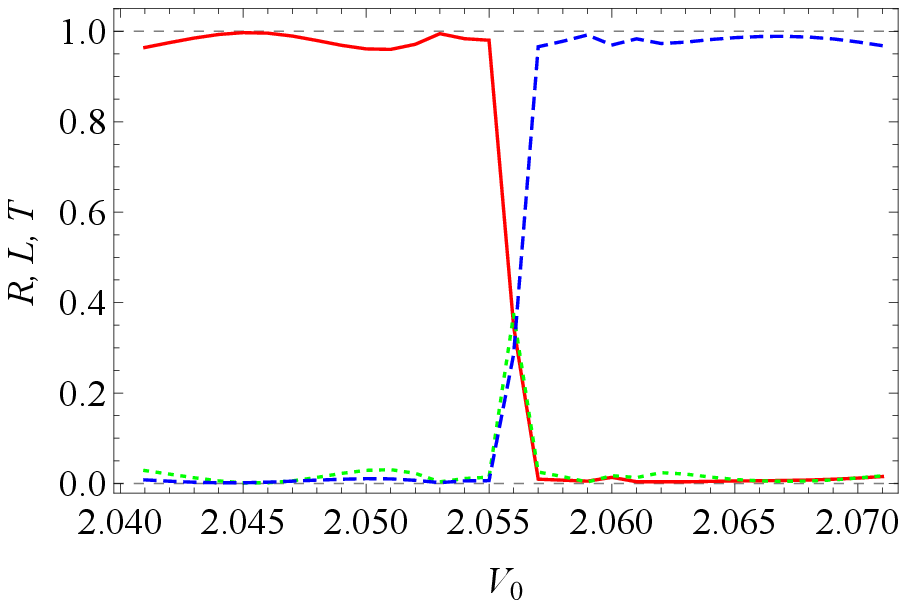}
    \begin{picture}(5,5)(5,5)
    \put(-28,132) {\color{black}{{\fcolorbox{white}{white}{\textbf{(d)}}}}}
    \end{picture}
    \caption{Scattering coefficients of (a) reflection,  $R$, (b) trapping,  $L$,  and (c) transmission, $T$,  for a flat-top soliton scattered by a wide reflectionless potential well with initial speed $v_0=0.1$. (d) A zoom-in of the three scattering coefficients for the left resonant scattering.  Parameters used: $u_0=0.093748$, $u_{0p}=0.2$, $\gamma=-0.998$, $\gamma_p=-0.9999999999999998$,  $g_1=0.5$, $g_2=1.0$, $g_{1p}=0.5$, $g_{2p}=1.0$.}
    \label{fig8}
\end{figure}

\begin{figure}[H]\centering
    \includegraphics[width=5cm,clip]{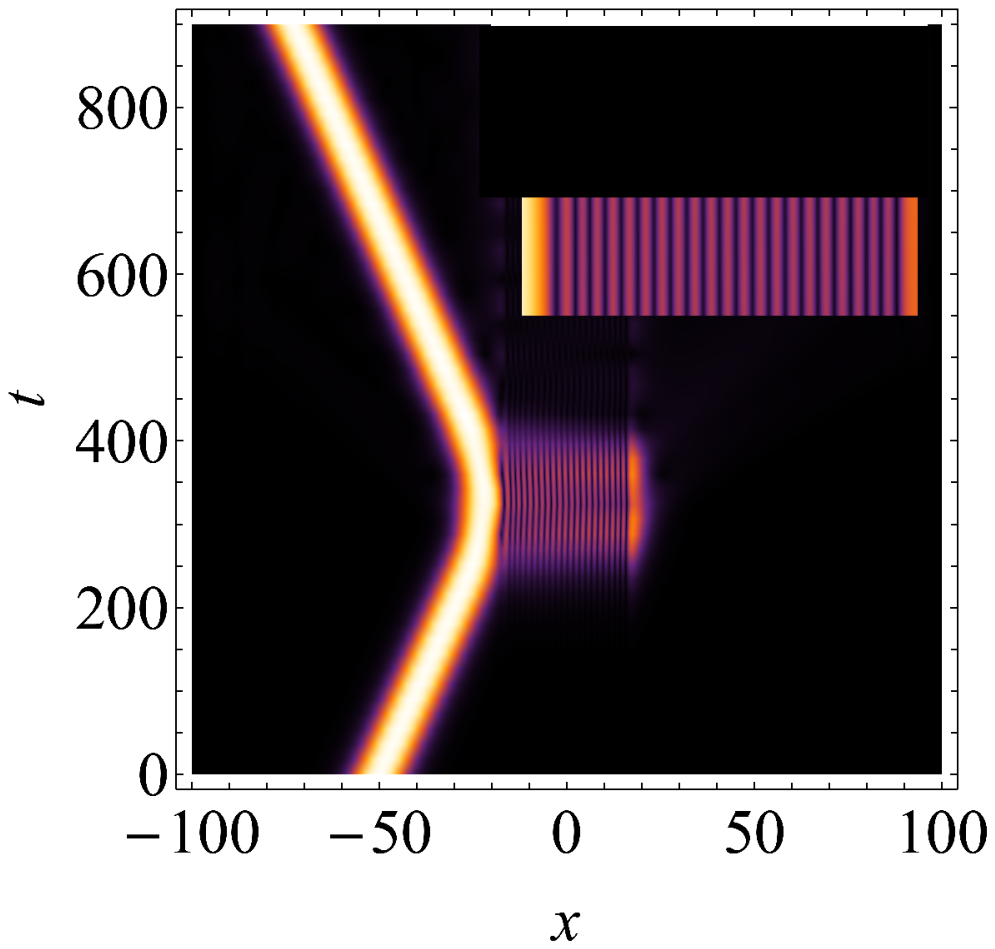}
    \begin{picture}(5,5)(5,5)
    \put(-28,40) {\color{black}{{\fcolorbox{white}{white}{\textbf{(a)}}}}}
    \end{picture}
    \includegraphics[width=5cm,clip]{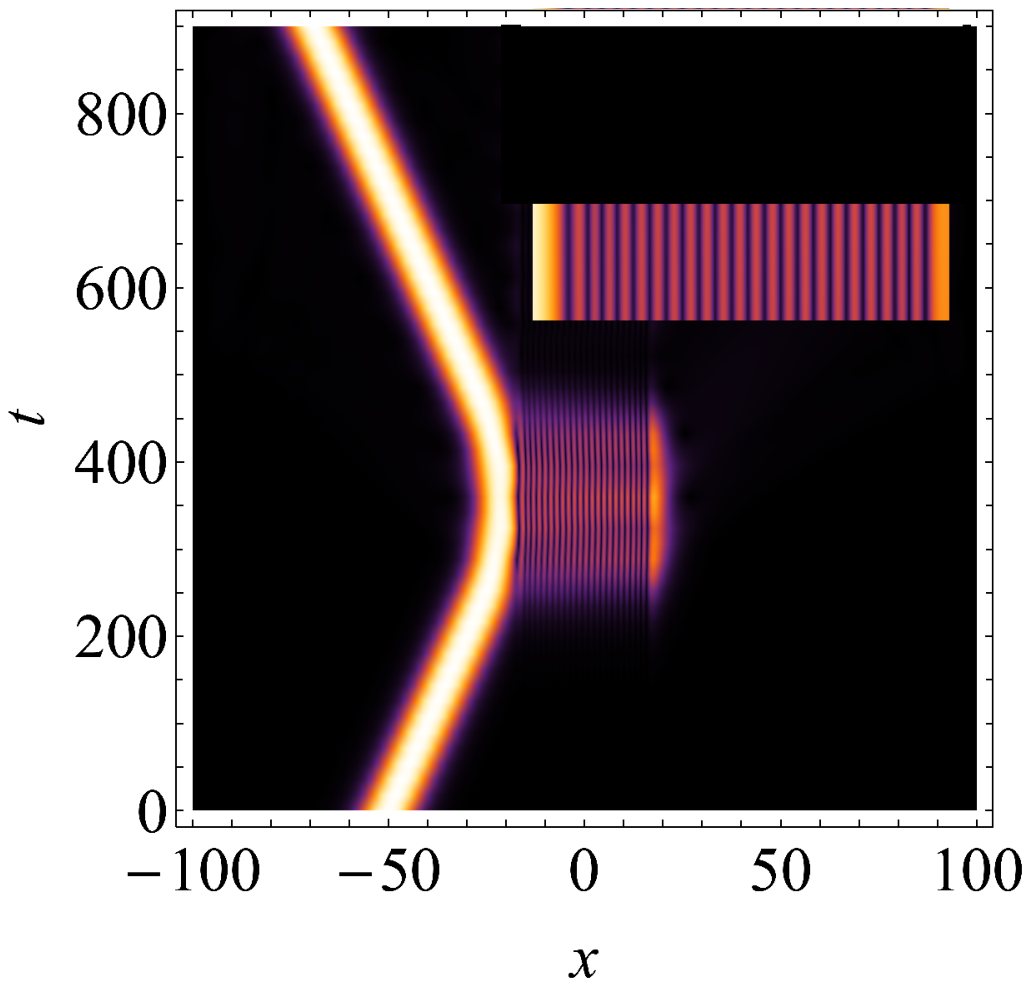}
    \begin{picture}(5,5)(5,5)
    \put(-28,40) {\color{black}{{\fcolorbox{white}{white}{\textbf{(b)}}}}}
    \end{picture}
    \includegraphics[width=5cm,clip]{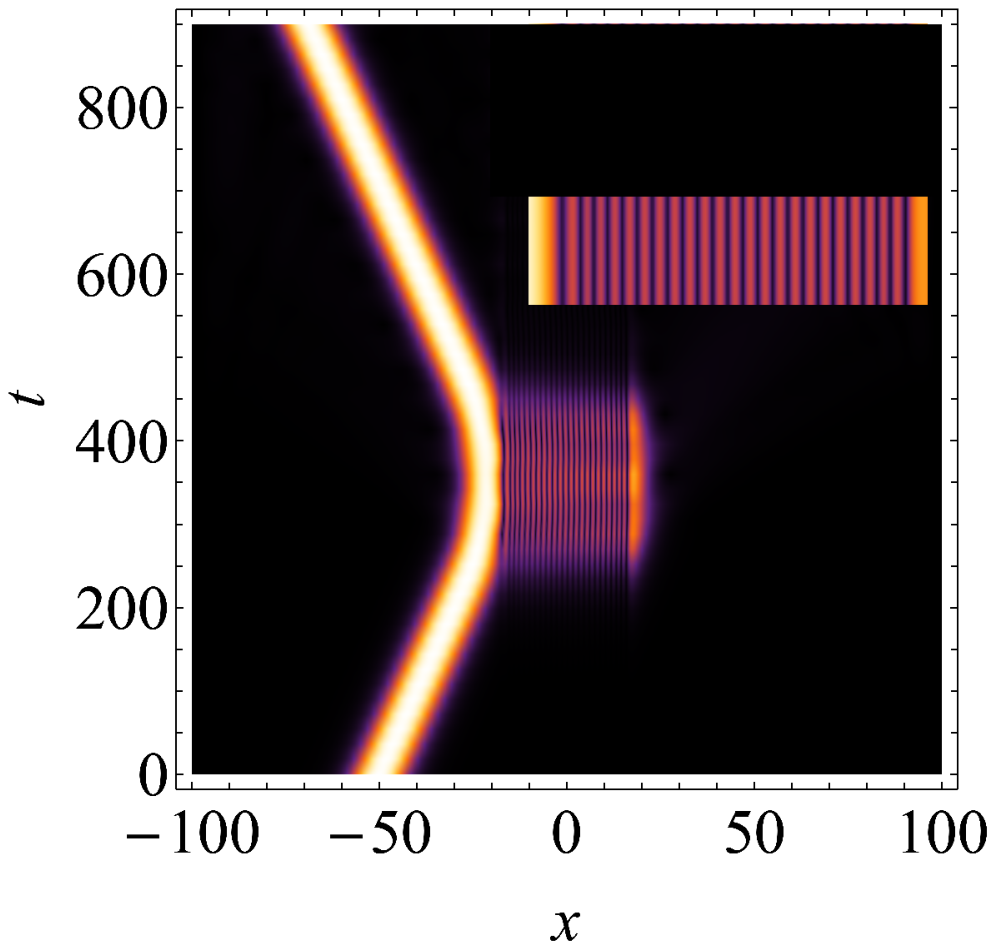}
    \begin{picture}(5,5)(5,5)
    \put(-28,40) {\color{black}{{\fcolorbox{white}{white}{\textbf{(c)}}}}}

    \end{picture}
    \vspace*{0.7cm}
    \caption{Full reflectance at resonances shown in Fig. \ref{fig8}(a)
    with a varying potential well depth (a) $V_0=2.046$ with a trapped mode
    of 22 nodes, (b) $V_0=2.236$ with a trapped mode of 23 nodes, and (c)
    $V_0=2.427$ with a trapped mode of 24 nodes.  Insets show a zoom-in of the
    multi-node trapped modes.  Other parameters:
    $u_0=0.093748$, $u_{0p}=0.2$, $\gamma=\gamma_p=-0.998$, $v_0=0.1$, $g_1=0.5$, $g_2=1.0$, $g_{1p}=0.5$, $g_{2p}=1.0$.}
    \label{fig9}
\end{figure}

For further investigation of the resonant modes, we have studied the
scattering of flat-top solitons by a wide reflectionless potential
with varying depth. In terms of the potential depth, $V_{0}$,
trapped modes with different numbers of nodes are excited. The
associated resonant scattering occurs always when a trapped mode is
excited. The number of nodes changes with the depth of the
potential. As $V_0$ is increased, a trapped mode with an increasing
number of nodes is found to occur at discrete values of $V_0$. This
behavior is due to a resonance between the fixed energy of the
incident soliton and the energy of the trapped modes. By increasing
the magnitude of $V_0$, the energy of trapped modes  increases and will
resonate one after the other with the energy of the incident
soliton. In Fig.~\ref{fig8}, we plot the transport coefficients,
reflectance $R=(1/n_0)\int_{-\infty}^{-l}|\psi(x,\tau)|^2dx$,
trapping $L=(1/n_0)\int_{-l}^{l}|\psi(x,\tau)|^2dx$, and
transmittance $T=(1/n_0)\int_{-l}^{\infty}|\psi(x,\tau)|^2dx$,
versus $V_0$. Here, $l$ is the length at which the potential is
negligible and $\tau$ is a time long after the scattering event. The
figure shows the multi-resonance behavior where three resonances are
observed. It should be noted that a similar critical behavior in the transport coefficients may also be obtained by fixing the potential depth and varying the initial speed of the incident soliton. However, changing the potential depth leads to exciting many multi-nodes trapped modes, which is indicated by the multi-resonance behavior shown in Fig.~\ref{fig8}. On the other hand, changing the soliton speed excites only one trapped mode at a certain critical speed.    
Figure \ref{fig9} shows the dynamics near resonances.

\section{Conclusions}
\label{concsec} From a generalized form of an exact solution to the
NLSE with dual nonlinearity, we have shown that a spectrum of
solutions exists containing four fundamentally different solutions,
namely bright solitons, kink solitons, flat-top solitons, and thin-top
solitons. We have shown that the four solutions are connected through
one parameter that transfers one solution to the other.

Using a perturbative expansion with a trial solution corresponding
to the scattered soliton, a general form of the reflectionless
potential is obtained in terms of one of the solutions to the NLSE
or any of its other integrable versions. Consequently, a spectrum of
potentials is also obtained in terms of a single parameter.

A series of numerical simulations to the scattering of flat-top and
thin-top solitons confirmed the reflectionless property of the
potential. Resonant scattering and quantum reflection were generated
in a similar manner as for the bright soliton scattering by a
reflectionless P\"oschl-Teller potential.

\end{document}